\documentclass[twocolumn,prd,aps,superscriptaddress]{revtex4-1}
\usepackage{graphicx}
\usepackage{appendix}
\usepackage{bm}
\usepackage{color}
\usepackage{slashed}
\usepackage{amsmath}
\usepackage{amsfonts}
\usepackage{verbatim}
\usepackage{amssymb}
\usepackage{graphicx}
\usepackage{epstopdf}
\usepackage{mathrsfs}
\usepackage{epsfig}
\usepackage{slashed}
\usepackage{bbold}%Matriz identidade
\usepackage{hyperref}
\usepackage{orcidlink}

\providecommand{\U}[1]{\protect\rule{.1in}{.1in}}
\begin{document}

\title{Finite temperature Casimir effect for a spinor field in cosmic dispiration spacetime}

\author{Jo\'{a}s Ven\^{a}ncio \orcidlink{0000-0003-0356-9983}}
\email[]{joasvenancioufpe@gmail.com}
\affiliation{Departamento de F\'{\i}sica, \href{https://www.ufpe.br/}{Universidade Federal de Pernambuco}, Recife, Pernambuco  50740-560, Brazil.}
\author{Herondy Mota \orcidlink{0000-0002-7470-1550}}
\email[]{hmota@fisica.ufpb.br}
\affiliation{Departamento de F\'{i}sica, \href{https://www.ufpb.br/}{Universidade Federal da Para\'{i}ba}, Jo\~{a}o Pessoa, Caixa Postal 5008, Brazil.}
\author{Azadeh Mohammadi \orcidlink{0000-0001-5720-7086}}
\email[]{azadeh.mohammadi@ufpe.br}
%\affiliation{Departamento de F\'{\i}sica, Universidade Federal de Pernambuco, Recife, Pernambuco  50740-560, Brazil.}
\affiliation{Departamento de F\'{\i}sica,  \href{https://www.ufpe.br/}{Universidade Federal de Pernambuco}, Recife, Pernambuco  50740-560, Brazil.}
%\affiliation{Departamento de F\'{i}sica, Universidade Federal da Para\'{i}ba, Jo\~{a}o Pessoa, Caixa Postal 5008, Brazil.}

\begin{abstract}

This study explores the finite temperature Casimir effect for a massive spinor field in cosmic dispiration spacetime, formed by the combination of a cosmic string and a screw dislocation using the generalized zeta function regularization method. First, we examine the cosmic string spacetime with a quasi-antiperiodic boundary condition, where the Casimir energy and its corrections depend on two nonzero heat kernel coefficients, one associated with the Euclidean divergence and the other with the nontrivial topology, both vanishing when renormalized. Interestingly, for specific choice of parameters the quasi-antiperiodicity effect can entirely cancel out the topological contribution, leaving only the Euclidean divergence. We then extend this analysis to cosmic dispiration spacetime. This configuration alters the spacetime topology, modifying the structure of the heat kernel coefficient related to the new nontrivial topology. In this case, the renormalized Casimir energy density can take positive or negative values and decreases exponentially as the field mass increases. Additionally, we examine the asymptotic behavior of the renormalized temperature correction term in the massless regime, showing that the spinor vacuum free energy vanishes at very high temperatures. At very low temperatures, it is dominated by the zero-temperature Casimir energy density.

\end{abstract}

\maketitle

\vspace{3cm}
\section{Introduction}

In his seminal 1948 paper \cite{Casimir1948}, Casimir provided the first genuine theoretical calculation of a purely quantum phenomenon that forms the foundation of modern understanding of the vacuum energy of fluctuating quantum fields, the celebrated Casimir effect \cite{Klimchitskaya2009, Milton2001}. He predicted the existence of a macroscopically observable attractive force between neutral metal plates caused by changes in vacuum fluctuations of the electromagnetic field. The Casimir result has been measured experimentally with high accuracy and agrees with his theoretical prediction \cite{Bressi2002, Lamoreaux1998, Mohideen1998}. Moreover, a few years after its discovery, the influence of finite-temperature effects was examined, leading to thermal corrections to the Casimir energy \cite{Fierz1960, Mehra1967, Brown1969}.
Driven by advances in particle and field theories, Casimir-like effects have been explored for fields with arbitrary spin in progressively more complex configurations \cite{Stokes2015, Muniz2018, Cunha2016, Mobassem2014, Bytsenko2005, Pereira2017, Chernodub2018}. %In our previous work \cite{Joas2024}, the thermal Casimir effect was studied for a massive spinor field defined on a flat spacetime with a circularly compactified spatial dimension. This identification condition induces a boundary condition that distorts the vacuum fluctuations of the spinor field, such as a material boundary does, producing the Casimir effect.

In our previous work \cite{Joas2024}, for instance, we explored the thermal Casimir effect for a massive spinor field in a flat spacetime with a nontrivial topology induced by an identification condition.
Interestingly, identification conditions of flat spacetimes allow for the construction of various metrics with a particular type of quasiregular singularities known as the conical singularities \cite{Tod1994}. These singularities often appear in spacetimes around a cosmic string, where the metric is locally flat and presents a deficit angle that modifies the usual $2\pi$ periodicity \cite{Hiscock1985, Vickers1987}. 
%Interestingly, identification conditions of flat spacetimes allow for the construction of various metrics with quasiregular singularities \cite{Tod1994}. A well-known example is the conical singularity that appears in spacetimes around a cosmic string, where the metric is locally flat and presents a deficit angle that modifies the usual $2\pi$ periodicity \cite{Hiscock1985, Vickers1987}.
The conical singularities of cosmic strings are described geometrically by a delta function-valued Riemann tensor.
Cosmic strings are geometric structures associated with linear topological defects, theorized to emerge from a symmetry breaking phase transition occurring after Planck time, in the first moments after the Big Bang \cite{Gibbons1990}. Quantum vacuum fluctuations of some physical observables induced by the conical structure of the cosmic string spacetime have been studied for the spinor field in Refs. \cite{Valdir2006, Aram2013, Belluci2014, Aram2017, Bezerra2024}. 

Another interesting kind of conical singularity appears in spacetimes around a chiral cosmic string, characterized geometrically by a delta function singularity in the Riemann and torsion tensors \cite{Hehl1976, Galtsov1993}.
The geometry associated with a chiral cosmic string possesses a helix-like structure analogous to dislocations in condensed matter physics. Specifically, in addition to the traditional disclination that is the cosmic string counterpart, spacetime incorporates a linear topological defect known as screw dislocation \cite{Puntigam1997}. This configuration can be considered as a linear topological defect carrying torsion and curvature \cite{Letelier1995}. The spacetime describing it referred to as cosmic dispiration spacetime, is elegantly formulated within the framework of the Einstein-Cartan theory of gravity \cite{Hehl1976}.  

%Cartan \cite{Cartan1966}, in turn, discovered the spinors a century ago as mathematical objects related to unknown representations of the $SO(N)$ group, the rotation group in $N$ dimensions.
Cartan proposed this theory soon after his prominent discovery of spinors a century ago \cite{Cartan1966}.
Spinors are objects that provide the least-dimensional faithful representations for the group $Spin(N)$, the universal covering of the group $SO(N)$, the rotation group in $N$ dimensions.
Spinor fields satisfying the standard Dirac equation describe spin-$1/2$ particles, such as protons, electrons, and neutrons. Since its formulation \cite{Dirac1928}, the Dirac equation and its separability properties have been studied in increasingly complex curved spacetimes \cite{Joas2017}. However, complete separability remains an open problem.
Interestingly, in cosmic dispiration spacetime, the Dirac equation is not only separable but also allows for analytical integration, which is important for studying the Casimir effect. 
%It is the most tractable spacetime after the cosmic string case and offers an interesting and largely unexplored setting for studying the Casimir effect, especially in the context of spinor fields.
The effect stems from the nontrivial topology of the cosmic dispiration spacetime, which modifies the quantum modes in the vacuum state, resulting in a nonzero vacuum energy \cite{Herondy2024}.  

Divergences are  an intrinsic feature of vacuum energy calculations within quantum field theory (QFT), and addressing them remains one of the most challenging aspects. A robust and sophisticated regularization method employs the generalized zeta function, defined through the eigenvalues of the differential operator that governs the dynamics of the relevant quantum field \cite{Elizalde1994, Hawking1977}. One of the main advantages of this method is its ability to incorporate temperature effects naturally through the Euclidean time formalism.  The divergent structure encoded in the generalized zeta function can be accessed from the asymptotic behavior of the heat kernel associated with the relevant operator \cite{Kac1966, Vassilevich2003}. 

While the Casimir effect in topologically nontrivial spacetimes has been widely studied, analyses specifically focusing on cosmic dispiration spacetime remain scarce. 
In Ref. \cite{Fernando1998}, the thermal Casimir effect for a scalar field was examined in the presence of a screw dislocation using the zeta function technique. More recently, Ref. \cite{Herondy2024} extended this investigation to cosmic dispiration spacetime in arbitrary dimensions. However, to our knowledge, no study has considered the spinor field case yet.

In this work, we explore the finite temperature Casimir effect using the generalized zeta function method for a massive spinor field defined on spacetime around a cosmic dispiration. This spacetime defect not only bridges cosmology and condensed matter physics but also stands as one of the most analytically tractable cases after the cosmic string, which emerges as a limiting scenario. The cosmic string case is analyzed from a broader perspective by imposing a quasi-antiperiodic boundary condition on the spinor field along the azimuthal direction. As a result, we generalize the results presented in {Ref.~\cite{Fursaev1997}} for spin-$1/2$ fields, where the authors studied the heat kernels of Laplace-type operators associated with spin-$s$ fields in spacetimes with conical singularities. The thorough examination of the zeta function method within the context of cosmic strings yields valuable results for analyzing divergent structures associated with nonzero heat kernel coefficients when screw dislocation effects are considered. Furthermore, it offers key insights that improve our understanding of the cosmic dispiration spacetime findings, representing an elegant spinorial generalization of those reported in {Ref.~\cite{Herondy2024}} for the scalar field.

The remainder of this work is organized as follows. Section \ref{S2} sets the scene for the introduction of spinors in curved spacetime, outlining the conventions adopted throughout the article. In Sec. \ref{S3}, we briefly discuss the intricate aspects of the generalized zeta function regularization method in the spinor context and explore its connection to the spinor vacuum free energy.
In Sec. \ref{S4}, we consider a massive spinor field in the background of cosmic dispiration spacetime and construct the complete set of normalized spinor solutions. Section \ref{S5} is divided into two parts. The first focuses on deriving the spinor heat kernel function, Casimir energy density, and temperature corrections in cosmic string spacetime. The second extends these analyses to cosmic dispiration, except for the temperature correction addressed in Sec. \ref{S6}. Finally, Sec. \ref{Conclusion} provides a summary of the paper. 
As an addendum, Appendix \ref{Heat kernel} presents an important integral representation of the components of the spinor heat kernel, which is utilized in calculating the Casimir energy and thermal corrections. Throughout this paper, we adopt the natural units where $c = G = \hbar = k_B = 1$.

\section{Preliminary assumptions}\label{S2}

%We can start with the path integral for spinor fields. 

Let $(\mathcal{M}, \boldsymbol{g})$ be an $N$-dimensional spin manifold equipped with a Euclidean metric $\boldsymbol{g}$ and a torsion-free connection $\nabla$ that is compatible with the metric. In a local coordinate system $\{x^{\mu}\} = \{t, x^{i}\}$ on $(\mathcal{M}, \boldsymbol{g})$, the line element is assumed to take the form
%For our purposes, it is especially useful to adopt a local coordinate system $\{x^{\mu}\} = \{t, x^{i}\}$ on $(\mathcal{M}, \boldsymbol{g})$, where $i = 1, 2, 3, \ldots, D = N-1$. In this coordinate system, the line element can be written in the following form:
\begin{align}\label{Generic Line Element}
    ds^2 = dt^2 + g_{ij}dx^{i} dx^{j} , 
\end{align}
where ${i, j =1, 2, 3, \ldots, D= N-1}$.
Here $g_{\mu\nu} = \boldsymbol{g}(\partial_{\mu}, \partial_{\nu})$ are the metric components in the coordinate frame $\{\partial_{\mu}\}$. 
%As usual, the imaginary time coordinate $t$ is compactified into a circumference length equal to the inverse of the temperature $\beta$, so that $(\mathcal{M}, \boldsymbol{g})$ is closed in the $t$ direction. 

To describe spinor fields, it is convenient to introduce an orthonormal frame of vector fields $\{\boldsymbol{X}_{\alpha}\}$, with respect to which the metric components can be written as
%The metric components concerning an orthonormal non-coordinate frame of vector fields $\{\boldsymbol{X}_{\alpha}\}$ defined on $(\mathcal{M}, \boldsymbol{g})$ are 
\begin{align}
   \boldsymbol{g}(\boldsymbol{X}_{\alpha}, \boldsymbol{X}_{\beta}) =  \delta_{\alpha\beta} , \quad \alpha,\beta = 1, 2, 3, \ldots, N,
\end{align}  
where $\delta_{\alpha\beta}$ is the Kronecker delta. 

Associated with the frame $\{\boldsymbol{X}_{\alpha}\}$ is its dual coframe of 1-forms, $\{\boldsymbol{e}^{\alpha}\}$, defined by the condition
\begin{align}
    \boldsymbol{e}^{\alpha}(\boldsymbol{X}_{\beta}) = \delta^{\alpha}_{\ \beta}.
\end{align}
Using this coframe, the line element \eqref{Generic Line Element} can be recast into a simple form as follows
%A convenient choice of the non-coordinate frame allows us to work with the metric more effectively, with 
\begin{align}\label{Constant metric component}
    ds^2 =  \boldsymbol{g}(\boldsymbol{X}_{\alpha}, \boldsymbol{X}_{\beta}) \,\boldsymbol{e}^{\alpha}\boldsymbol{e}^{\beta} .
   \end{align}

The covariant derivative $\nabla_{\alpha}$ along $\boldsymbol{X}_{\alpha}$ is
determined by its action on the basis vectors
\begin{align}\label{Spin connection}
\nabla_{\alpha}\boldsymbol{X}_{\beta} = \omega_{\alpha\beta}^{\phantom{\alpha\beta}\varepsilon}\boldsymbol{X}_{\varepsilon} ,
\end{align}
which introduces the spin connection components $\omega_{\alpha\beta}^{\phantom{\alpha\beta}\gamma}$. 
These components are antisymmetric in their last two indices when all indices are lowered, as expressed by ${\omega_{\alpha\beta\varepsilon} = \omega_{\alpha\beta}^{\phantom{\alpha\beta}\eta}\delta_{\eta\gamma} = -\omega_{\alpha\gamma\beta}}$, a property that follows from the metric compatibility condition $\nabla \boldsymbol{g} = 0$. 
It should be noted that the index of $\omega_{\alpha\beta}^{\phantom{\alpha\beta}\gamma}$ can be lowered using $\delta_{\alpha\beta}$ and raised with $\delta^{\alpha\beta}$, allowing the frame indices to be manipulated without restriction.

%Since $\boldsymbol{g}$ is a covariantly constant tensor, $\omega_{\alpha\beta}^{\phantom{\alpha\beta}\varepsilon}$ with all lower indices is antisymmetric in its last two indices, satisfying ${\omega_{\alpha\beta\varepsilon} = \omega_{\alpha\beta}^{\phantom{\alpha\beta}\eta}\delta_{\eta\varepsilon} = -\omega_{\alpha\varepsilon\beta}}$. Note that the spin connection components have their indices raised using $\delta^{\alpha\beta}$ and lowered using $\delta_{\alpha\beta}$, allowing the frame indices to be manipulated without restriction. Specifically, $\omega_{\alpha}^{\phantom{\alpha}\beta\varepsilon}$ satisfies ${\omega_{\alpha}^{\phantom{\alpha}\beta\varepsilon} = \omega_{\alpha}^{\phantom{\alpha}[\beta\varepsilon]}}$, where the indices in square brackets indicate antisymmetrization. 

In $N$ dimensions, the spinor space has dimension $2^{[N/2]}$, where $[N/2]$ denotes the floor of $N/2$. Consequently, the orthonormal frame $\{\boldsymbol{X}_{\alpha}\}$ can be faithfully represented by a set of $2^{[N/2]} \times 2^{[N/2]}$ Dirac matrices $\{\gamma_{\alpha}\}$, which satisfy the Clifford algebra associated with $(\mathcal{M}, \boldsymbol{g})$
%In $N$ dimensions, the orthonormal frame $\{\boldsymbol{X}_{\alpha}\}$ can be faithfully represented by $4 \times 4$ Dirac matrices $\{\gamma_{\alpha}\}$ satisfying the Clifford algebra over $\mathcal{M}$
\begin{align}\label{Clifford algebra}
     \gamma_{\alpha} \gamma_{\beta} + \gamma_{\beta} \gamma_{\alpha} = 2 \, \boldsymbol{g}(\boldsymbol{X}_{\alpha}, \boldsymbol{X}_{\beta}) \mathbb{1} ,
\end{align}
with $\mathbb{1}$ standing for the $2^{[N/2]} \times 2^{[N/2]}$ identity matrix in spinor space.
With this convention, the Dirac matrices are hermitian, expressed as $\gamma_{\alpha}^{\dagger} = \gamma_{\alpha}$. 
The spinor fields are then represented by column vectors on which these matrices act.

The standard Euclidean Dirac operator on $(\mathcal{M}, \boldsymbol{g})$ is given by
\begin{align}
    \slashed{D} = i\gamma^{\alpha} \nabla_{\alpha} ,
\end{align}
where $\nabla_{\alpha}$ is the spinor covariant derivative 
\begin{align}
\nabla_{\alpha} = \boldsymbol{X}_{\alpha} - \frac{1}{4}\,\omega_{\alpha}^{\phantom{\alpha}\beta\varepsilon}\gamma_{\beta}\gamma_{\varepsilon} .
\end{align}
%$\partial_{\alpha}$ denotes the partial derivative along the vector field $\boldsymbol{X}_{\alpha}$.
%The dynamic of a Dirac field $\Psi$ is governed by the Euclidean action functional 
%\begin{align}\label{Euclidean action}
%    \mathcal{S}_E(\Psi, \bar{\Psi}) = 
%\int_{0}^{\beta} dt\int d^{3} x\, \,\mathcal{L}(\Psi, \bar{\Psi}) ,
%\end{align}
%where the lagrangian density $\mathcal{L}$ can be taken as
%\begin{align}
%  \mathcal{L}(\Psi, \bar{\Psi}) = \frac{\sqrt{g} }{2}\left(\bar{\Psi}\slashed{D}\Psi - \overline{\slashed{D}\Psi}\Psi \right) + m \bar{\Psi}\Psi % .
%\end{align}
%with $g$ standing for the metric determinant.
This operator is defined to act appropriately on a Dirac field $\boldsymbol{\Psi}(x)$, whose dynamics is described by the Euclidean action functional
\begin{align}\label{Euclidean action}
    \mathcal{S}_E(\boldsymbol{\Psi}, \boldsymbol{\bar{\Psi}}) = 
\int_{0}^{\beta} dt\int d^{D} x\,\sqrt{g}\,\boldsymbol{\bar{\Psi}}(x)\slashed{D}(m)\boldsymbol{{\Psi}}(x) ,
\end{align}
where $g$ stands for the metric determinant and $\slashed{D}(m)$ the Dirac operator with the mass term
\begin{align}
    \slashed{D}(m) = \slashed{D} + i m . 
\end{align}
The spinor field $\boldsymbol{\Psi}$ is a Grassmann-valued field, meaning that it assigns a Grassmann variable to each point $x$. The same applies to the conjugate spinor $\boldsymbol{\bar{\Psi}}$, which, in Euclidean signature, is simply the Hermitian conjugate of $\boldsymbol{\Psi}$, i.e., ${\boldsymbol{\bar{\Psi}} = \boldsymbol{\Psi}^{\dagger}}$. Therefore, there is no need to differentiate between them.
%However, when the four-component spinors $\{\boldsymbol{\psi}_j\}$ are used as a basis for expanding the spinor fields $\boldsymbol{\Psi}$ and $\boldsymbol{\bar{\Psi}}$, it is convenient to represent them as:
However, when the spinor eigenmodes $\boldsymbol{\psi}_j$ of $\slashed{D}$, corresponding to the eigenvalues $\lambda_j$, are employed as a basis for expanding the spinor fields $\boldsymbol{\Psi}$ and $\boldsymbol{\bar{\Psi}}$, it is convenient to express them as:
\begin{align} 
\boldsymbol{{\Psi}}(x) = \sum_j \Psi_j \boldsymbol{\psi}_j(x), \label{Field expansion Psi}\\ 
\boldsymbol{\bar{\Psi}}(x) = \sum_j \bar{\Psi}_j \boldsymbol{\psi}^{\dagger}_j(x). \label{Field expansion Psi bar} \end{align}
The independent coefficients $\Psi_j$ and $\bar{\Psi}_j$ characterize the Grassmannian nature of the spinor field and are not connected by complex or Hermitian conjugation. %Notice that the spinor conjugate operation, in addition to transposing and complex conjugating the spinor $\boldsymbol{\psi}_j$ also converts any ${\Psi}_j$  into $\bar{\Psi}_j $. 
In general, the index $j$ that labels the field modes comprises both discrete and continuous quantum numbers.

In our previous work \cite{Joas2024}, we provided a detailed derivation of the one-loop partition function $Z$, associated with a Dirac field, for a canonical ensemble at temperature ${T = \beta^{-1}}$ in flat spacetime. By following the same steps, one can demonstrate that its validity extends seamlessly to the domain of curved spacetimes as well. Specifically, we have
\begin{align}\label{Divergent det}
Z &= \int  \mathcal{D}\bar{\Psi} \mathcal{D}\Psi \, e^{\mathcal{S}_{E}(\Psi, \bar{\Psi})} \nonumber\\
& = \prod_{j} \dfrac{\lambda_j + i m}{\mu}= \text{det}\left[ \dfrac{\slashed{D}(m)}{\mu}  \right] ,
\end{align}
where $\mu$ is a scale parameter with dimensions of inverse length or mass. For a more detailed discussion of the parameter $\mu$, refer to \cite{Hawking1977, Elizalde1990}.

The divergence arising from the infinite product of eigenvalues requires a proper regularization procedure. In this paper, we employ an efficient and elegant method based on the generalized zeta function, which corresponds to the zeta function of an operator.

\section{Generalized zeta and heat kernel functions of the Dirac operator}\label{S3}

Let $L$ be a non-negative, self-adjoint, second-order elliptic differential operator. A widely used regularization method for the one-loop functional determinant of $L$ relies on the generalized zeta function, which is defined as
\begin{equation}\label{generalized zeta}
    \zeta_{L}(w) = \sum_j \lambda_{j}^{-w} .
\end{equation}
In this expression, $\lambda_j$ denotes the eigenvalues of $L$. Although the summation symbol is used, the spectrum of $L$ does not need to be discrete; the index 
$j$ simply labels the field modes. In $N$ dimensions, the series \eqref{generalized zeta} converges for $\text{Re}(w) > N/2$ and remains regular at $w = 0$. Moreover, it admits an analytic continuation to other values of $w$ \cite{Seeley1967}.
 
The corresponding regularized determinant of $L$ can be expressed in terms of the generalized zeta function as
\begin{equation}\label{regulated determinant}
    \text{det}(L) = e^{-\zeta_{L}'(0)},
\end{equation}
where $\zeta_{L}'(w)$ denotes the derivative of $\zeta_{L}(w)$ with respect to $w$. This formulation follows from the identity ${\text{ln} \lambda_j = -\frac{d}{dw} \lambda_j^{-w}\big|_{w=0}}$ for non-negative eigenvalues $\lambda_{j}$. As a result, one recovers the familiar relation  $e^{\sum_{j}\text{ln}\lambda_j} = \prod_j \lambda_j$.

Since the spectrum of $\slashed{D}$ includes negative eigenvalues, the regularized determinant defined in \eqref{regulated determinant} cannot be directly applied to the Dirac operator. Therefore, it is necessary to establish a suitable definition for its determinant. To this end, we begin by considering the massless case. Let the positive eigenvalues of $\slashed{D}$ be denoted by $\lambda_j$, and the negative ones by $-\nu_j$. The associated zeta function can then be written as the sum of two separate contributions given by
\begin{align}
    \zeta_{\slashed{D}}(w)  =  \sum_{j} \lambda_{j}^{-w}  +  (-1)^{-w}\sum_{j} \nu_{j}^{-w} . 
\end{align}
Moreover, this expression can be decomposed into a symmetric and an antisymmetric part under the transformation $\slashed{D} \rightarrow -\slashed{D}$, as follows
%\begin{align}\label{Zeta and eta}
%    &\zeta_{\slashed{D}}(w)   =  \sum_{j} \left(\frac{\lambda_{j}^{-w} + \nu_{j}^{-w}}{2} + \frac{\lambda_{j}^{-w} - \nu_{j}^{-w}}{2}  \right)  \nonumber\\
%    &+  (-1)^{-w}\sum_{j}\left(\frac{\lambda_{j}^{-w} + \nu_{j}^{-w}}{2} - \frac{\lambda_{j}^{-w} - \nu_{j}^{-w}}{2}  \right)\nonumber\\
 %   &= \frac{1}{2}\left[1+ (-1)^{-w} \right]   \zeta_{\slashed{D}^2}(w/2) +  \frac{1}{2}\left[1- (-1)^{-w} \right]\eta_{\slashed{D}}(w) .
%\end{align}
\begin{align}\label{Zeta and eta}
    \zeta_{\slashed{D}}(w)  & = \frac{1}{2}\left[1+ (-1)^{-w} \right]  \left(\sum_{j} \lambda_{j}^{-w} + \sum_{j} \nu_{j}^{-w} \right)\nonumber\\
    &+  \frac{1}{2}\left[1- (-1)^{-w} \right]\left(\sum_{j} \lambda_{j}^{-w} -\sum_{j} \nu_{j}^{-w} \right).
\end{align}
The symmetric part is represented by $\zeta_{\slashed{D}^2}(w/2)$, where $\slashed{D}^2$ is the spinor Laplacian on $(\mathcal{M}, \boldsymbol{g})$, 
\begin{align}
    \slashed{D}^2 = -(\gamma^{\alpha}\nabla_{\alpha})^2,
\end{align}
which is a non-negative, self-adjoint, second-order Laplace-like operator. The antisymmetric part is a spectral function given by the expression
\begin{align}
    \eta_{\slashed{D}}(w)  & =  \sum_{j} \lambda_{j}^{-w}  -  \sum_{j} \nu_{j}^{-w} ,
\end{align}
which measures the spectral asymmetry of $\slashed{D}$ \cite{Vassilevich2011}. 

In some cases, it is more convenient to express the functions $\zeta_{\slashed{D}^2}(w/2)$ and $\eta_{\slashed{D}}(w)$ in terms of integral representations derived through a Mellin transform \cite{Vassilevich2011}. For instance, the integral representation for $\zeta_{\slashed{D}^2}(w/2)$ is as follows
%Both $\zeta_{\slashed{D}^2}(w/2)$ and $\eta_{\slashed{D}}(w)$ can also be expressed through integral representations obtained via a Mellin transform \cite{Vassilevich2011}. For example, the integral representation for $\zeta_{\slashed{D}^2}(w/2)$ is given by:
\begin{align}\label{integral representation}
   \zeta_{\slashed{D}^2}(w)  = \frac{1}{\Gamma(z)} \int_{0}^{\infty} d\tau\,\tau^{w-1} K_{\slashed{D}^2}(\tau) ,
\end{align}
where $K_{\slashed{D}^2}(\tau)$ represents the heat kernel function, defined as
\begin{align}\label{heat kernel trace}
    K_{\slashed{D}^2}(\tau) = {\text{Tr}}\left(e^{-\tau \slashed{D}^2} \right) = \sum_j e^{-\tau \lambda_j^2} .
\end{align}
Here, $\text{Tr}$ denotes the trace operation, which includes contributions from the spinor degrees of freedom.
The expression above diverges as ${\tau \rightarrow 0}$, and in this limit, the heat kernel function admits the following asymptotic expansion
\begin{align}\label{heat equation expansion} 
K_{\slashed{D}^2}(\tau) \sim \frac{1}{(4 \pi \tau)^{N/2}} \sum_{p} c_{p}(\slashed{D}^2) \,\tau^{p}, \,\, p = 0, \frac{1}{2}, 1, \frac{3}{2}, \cdots. 
\end{align} 
Thus, the heat kernel coefficients $c_{p}(\slashed{D}^2)$ not only characterize the divergences of the zeta function but also play a central role in determining the spinor vacuum free energy $F$, defined as \cite{Landau1980}
\begin{equation}\label{free energy} F(\beta, m, \mu) = -\frac{1}{\beta}\, \text{ln} Z, 
\end{equation} 
which inherently includes divergent contributions, as will become evident in the subsequent discussion. 

The zeta function \eqref{Zeta and eta} can be extended to the entire complex plane as a meromorphic function, well-defined apart from the sign ambiguity in $(-1)^{-w} = e^{\pm i \pi w}$. Since both functions $\zeta_{\slashed{D}^2}(w/2)$ and $\eta_{\slashed{D}}(w)$ remain regular at $w=0$, it follows from \eqref{regulated determinant} that the Dirac determinant can be expressed in the form
%\begin{align}
 %   \text{det} \slashed{D}  =  ( \text{det}\slashed{D}^{2})^{1/2}\,e^{i \phi} ,
%\end{align}
\begin{align}
    \text{ln\,det}(\slashed{D})  &= \frac{1}{2}\left[-\zeta_{\slashed{D}^2}(0) \pm  i\pi \zeta_{\slashed{D}^2}(0) \mp i\pi\eta_{\slashed{D}}(0) \right] \label{detD}\\
     &= \frac{1}{2}\,\text{ln\,det}(\slashed{D}^{2}) + a(\slashed{D}) \label{Massless anomaly},
\end{align}
where the term $a(\slashed{D})$ is given by 
\begin{align}\label{Phase}
     a(\slashed{D}) = \frac{i\pi}{2}[\pm  \zeta_{\slashed{D}^2}(0) \mp \eta_{\slashed{D}}(0)].
\end{align}
With this, the generalized zeta function \eqref{Zeta and eta}, together with \eqref{detD}, allows  us to express the one-loop partition function \eqref{Divergent det} in the massless limit as follows
\begin{equation}\label{Log Z 1}
\text{ln}Z = -\frac{1}{2} \left[\zeta_{\slashed{D}^2}'(0) + \text{ln}(\mu^2) \zeta_{\slashed{D}^2}(0)\right] + a(\slashed{D}) .
\end{equation}

Although the $\pm \zeta_{\slashed{D}^2}(0)$ term can be eliminated by an appropriate choice of the scale parameter $\mu$, the $\pm$ sign preceding $\eta_{\slashed{D}}(0)$ introduces a significant ambiguity in the definition of the determinant and consequently, in the vacuum free energy $F$. %, defined as \cite{Landau1980}: 
%\begin{equation}\label{free energy} F = -\frac{1}{\beta}\, \text{ln} Z. 
%\end{equation} 
However, in the even dimensional case ${N = 2d}$, the hermitian chiral matrix $\gamma^{N+1}$ anti-commutes with the Dirac operator
\begin{align}
    \gamma^{N+1} \slashed{D} = -  \slashed{D} \gamma^{N+1} . 
\end{align}
This implies that if $\boldsymbol{\psi}_j$ is an eigenmode of $\slashed{D}$ with a non-zero eigenvalue $\lambda_j$, then $\gamma^{N+1}\boldsymbol{\psi}_j$ is also an eigenmode, corresponding to the eigenvalue $-\lambda_j$. Therefore, the nonzero spectrum of the massless operator $\slashed{D}$ is symmetric, which implies that the eta function $\eta_{\slashed{D}}(0)$ vanishes identically. As a result, any information on the sign ambiguity in $F$ is removed entirely. In the odd-dimensional case $N = 2d + 1$, the asymmetry of $\slashed{D}$'s spectrum and the nontrivial value of $\eta_{\slashed{D}}(0)$ result in a sign ambiguity in $F$, which presents an undetermined overall sign when $\eta_{\slashed{D}}(0) \neq 0$.

The expression \eqref{Massless anomaly} provides some insight into the so-called multiplicative anomaly appearing 
in the massive case \cite{Kassel1989}, which refers to the fact that zeta-function regularized determinants do not satisfy the standard relation
\begin{align}
    \text{ln det}(A B)  = \text{ln det}(A) + \text{ln det}(B),
\end{align}
where $A$ and $B$ denote two elliptic differential operators. The massive extension of the expressions \eqref{detD} and \eqref{Massless anomaly} presents subtle complexities, which are thoroughly addressed in Ref. \cite{Zerbini2003}. For even dimensions, $N = 2d$, the resulting expression takes the form
\begin{align}
    \text{ln\,det} [\slashed{D}(m)]  &= -\frac{1}{2}\zeta_{\slashed{D}^2(m)}(0) \pm  \frac{i\pi}{2}\zeta_{\slashed{D}^2(m)}(0) \nonumber\\
    &+  \sum_{j=1}^{d}\sum_{p=1}^{j} (-1)^{j} \frac{c_{d-j} m^{2j} }{j! (2p-1)} \label{detD(m)}\\
     &= \frac{1}{2}\,\text{ln\,det}[\slashed{D}^2(m)] + a(\slashed{D}(m)) \label{Massive anomaly},
\end{align}
where the anomalous term $a(\slashed{D}(m))$ is given by
\begin{align}\label{Phase}
     a(\slashed{D}(m)) = \pm \frac{i\pi}{2} \zeta_{\slashed{D}^2(m)}(0) + \sum_{j=1}^{d}\sum_{p=1}^{j} (-1)^{j} \frac{c_{d-j} m^{2j} }{j! (2p-1)},
\end{align}
%mencionar o caso impar
with the massive operator $\slashed{D}^2(m)$ defined as
\begin{align}
    \slashed{D}^2(m) := \slashed{D}^{\dagger}(m)\slashed{D}(m) = \slashed{D}^2 + m^2.  
\end{align}
The coefficients ${c_p=c_p(\slashed{D}^2})$ are the heat kernel coefficients associated with the asymptotic expansion of the heat kernel $ K_{\slashed{D}^2}(\tau)$.

As discussed previously, the term $\pm \zeta_{\slashed{D}^2(m)}(0)$ can be absorbed through a redefinition of the scale parameter $\mu$.
Furthermore, since the coefficients $c_p$ are accompanied by non-negative mass exponents and an infinitely heavy field is expected to exhibit no quantum fluctuations, the renormalization process effectively incorporates this contribution, consistent with the renormalization condition \eqref{renormalization condition}.
Consequently, the anomalous term does not physically affect the one-loop partition function for the massive case, which can be written as
\begin{equation}\label{Log Z 2}
\text{ln}Z = -\frac{1}{2} \left[\zeta_{\slashed{D}^{2}(m)}'(0) + \text{ln}(\mu^2)  \zeta_{\slashed{D}^{2}(m)}(0)\right] ,
\end{equation}
precisely matching the expression presented in our earlier work \cite{Joas2024}.

Therefore, the spinor fields $\boldsymbol{\psi}_j$ we are looking for satisfy the following equation
%Note in particular that the Dirac fields $\boldsymbol{\psi}_j$ are eigenfunctions of $\slashed{D}^{2}(m)$ with non-negative eigenvalues
\begin{align}\label{Eigenvalue equation fo D^2}
   \slashed{D}^{2}(m) \boldsymbol{\psi}_j = \left( \lambda_{j}^{2} + m^2  \right)\boldsymbol{\psi}_j .
\end{align}
This gives our approach a mixed character, as the fields $\boldsymbol{\psi}_j$ are eigenmodes of a Laplace-type operator with non-negative eigenvalues, similar to scalar fields, yet they satisfy anti-periodic conditions in the imaginary time $t$
\begin{align}\label{temporal periodicity}
    \boldsymbol{\psi}_j(t,\boldsymbol{r}) = -    \boldsymbol{\psi}_j(t+\beta,\boldsymbol{r}),
\end{align}
where ${\boldsymbol{r} = (x^1, x^2,\ldots, x^D)}$ denotes the set of $D$ spatial coordinates. This condition corresponds to compactifying the imaginary time coordinate $t$ into a circle of circumference equal to the inverse temperature $\beta$. This amounts to endowing $(\mathcal{M}, \boldsymbol{g})$ with the topology ${\mathbb{S}^1(\text{time}) \times M^{D}}$, where $M^{D}$ represents a $D$-dimensional base manifold whose line element is determined by the spatial component of \eqref{Generic Line Element}. Consequently, it becomes especially practical to express the operator $\slashed{D}^2(m)$ as
\begin{equation}\label{Laplace type operator}
        \slashed{D}^2(m) =  - \partial^{2}_{t} + \slashed{\nabla}^2(m) ,
\end{equation}
where $\slashed{\nabla}^2(m)$ denotes the massive Dirac operator defined on $M^D$, explicitly given by
\begin{align}
 \slashed{\nabla}^2(m) = -(\gamma^{i}\nabla_i)^2 + m^2. 
\end{align}
Moreover, as $\partial_t$ is an obvious Killing vector field of our metric, it follows that the spinor fields satisfying the condition \eqref{temporal periodicity} can be expressed as follows
\begin{equation}\label{time decomposition}
    \boldsymbol{\psi}_j(x) = e^{-i\omega_{n} t} \boldsymbol{\psi}_{a}(\boldsymbol{r}) , \quad  \omega_n = \frac{2\pi}{\beta}\left(n + \frac{1}{2}  \right) ,
\end{equation}
where the label $j$ now represents two quantum numbers, ${j = (n, a)}$, with ${n \in \mathbb{Z}}$. %This leads to the eigenvalues to be written as: 
As a result, the eigenvalues can be expressed as the sum of two components, given by
\begin{align}
    \lambda_j^2 =   \frac{4\pi^2}{\beta^2}\left(n + \frac{1}{2}  \right)^2 + \lambda_{a}^{2},
\end{align}
with $\lambda_{a}$ representing the spatial contribution to $\lambda_j$. 

This enables us to express the zeta function $\zeta_{\slashed{D}^2(m)}$ in a more convenient form, as detailed in \cite{Joas2024}. The final result is presented as follows
\begin{align}\label{Zeta of D(m) square}
   &{~}\zeta_{\slashed{D}^{2}(m)}(w)   =   \frac{\beta}{\sqrt{4\pi}} \left[ \frac{\Gamma\left(w- 1/2\right)}{\Gamma(w)} \,\zeta_{\slashed{\nabla}^2(m)}\left(w- 1/2\right)  \right. \nonumber\\
 &+ \left. \frac{2}{\Gamma(w)} \sum_{n=1}^{\infty} \int_{0}^{\infty} d\tau\,\tau^{w-\frac{3}{2}}  \cos(\pi n) e^{-\frac{\beta^2}{4 \tau}n^2}  K_{\slashed{\nabla}^2(m)}(\tau)\right]  ,
 \end{align}
where both $\zeta_{\slashed{\nabla}^2(m)}$ and $K_{\slashed{\nabla}^2(m)}$ are associated with the spinor operator $\slashed{\nabla}^2(m)$. Furthermore, the spinor vacuum free energy, derived from the partition function \eqref{Log Z 2}, can be decomposed into two distinct terms, outlined as  \cite{Joas2024}
\begin{align}\label{Reg free energy}
    F(\beta, m, \mu) =  E(m, \mu) + \Delta F(\beta, m). 
\end{align}
$E(m, \mu)$ represents the zero-temperature component of the vacuum free energy and is given by
\begin{align}\label{Effetive energy}
   E(m, \mu) = -\frac{1}{2}\,\text{FP}[\zeta_{\slashed{\nabla}^2(m)}(-1/2)] + \frac{ c_{\frac{N}{2}}(m)}{(4\pi)^{N/2}}\,\text{ln}\left(\frac{e \mu}{2}\right), 
\end{align}
where $\text{FP}[\zeta_{\slashed{\nabla}^2(m)}(-1/2)]$ stands for the finite part of $\zeta_{\slashed{\nabla}^2(m)}(-1/2)$, and the massive heat kernel coefficient $ c_{\frac{N}{2}}(m)$ is expressed as 
\begin{align}\label{c_2(m) heat kernel coefficient}
    c_{\frac{N}{2}}(m) = \sum_{p=0}^{N/2} \frac{(-1)^p}{p!}\, c_{\frac{N}{2}-p} m^{2 p} .
\end{align} 
The remaining $\beta$-dependent term, $\Delta F(\beta, m)$, represents the temperature corrections to the vacuum free energy and takes the form
\begin{align}\label{Generic temperature correction}
  \Delta F(\beta, m) & =  \frac{1}{\sqrt{4\pi}} \sum_{n= 1}^{\infty} \cos(\pi n) \times \nonumber\\
  & \times \int_{0}^{\infty} d\tau\,\tau^{-3/2}  e^{-\frac{\beta^2}{4 \tau}n^2} K_{\slashed{\nabla}^2(m)}(\tau) . 
\end{align}

Given that we are addressing a purely quantum phenomenon, the contribution arising from heat kernel coefficient $c_{\frac{N}{2}}(m) $, which grows with non-negative powers of the mass, is not expected to dominate in the large-mass limit. Consequently, this contribution must be subtracted during the renormalization process to derive a properly renormalized expression for the vacuum energy. Through this procedure, the resulting expression corresponds to the spinor Casimir energy at zero temperature
\begin{align}\label{Renormalized Casimir energy}
   E_{\text{C}}(m) = -\frac{1}{2}\,\text{FP}[\zeta_{\slashed{\nabla}^2(m)}(-1/2)], 
\end{align}
whose renormalized version satisfies the renormalization condition \cite{Bordag2000}
\begin{align}\label{renormalization condition}
    \underset{m \rightarrow \infty}{\text{lim}} E_{\text{C}}^{\text{ren}}(m) = 0 .
\end{align}
Notably, this eliminates any dependence of the vacuum energy on the scale factor $\mu$, thereby resolving the ambiguity inherent in the zeta function regularization prescription when $c_{\frac{N}{2}}(m) \neq 0$.
In the massless case (${m = 0}$), it is important to highlight that all the ambiguity resides entirely in the coefficient $c_{\frac{N}{2}}$. Consequently, such ambiguity arises only if ${c_{\frac{N}{2}} \neq 0}$.

We reduce the challenge of determining the spinor Casimir energy and its thermal corrections to evaluating the global heat kernel function of the operator $\slashed{\nabla}^2(m)$ on $M^D$, which can be written as
\begin{align}\label{Global heat kernel}
   K_{\slashed{\nabla}^2(m)}(\tau) = \int d^{D}\boldsymbol{r}\,\sqrt{g^{(D)}}\,\text{tr}\left[\boldsymbol{K}_{\slashed{\nabla}^2(m)}(\boldsymbol{r}, \boldsymbol{r}, \tau)\right], 
\end{align}
where $g^{(D)}$ is the spatial part of the metric determinant. The operation tr represents the trace over the spinor indices of
$\boldsymbol{K}_{\slashed{\nabla}^2(m)}(\boldsymbol{r}, \boldsymbol{r}', \tau)$, a two-point function locally defined as the solution to the 
 Schr\"{o}dinger-like equation
\begin{equation}\label{heat equation} 
\left[\frac{\partial}{\partial \tau} + \slashed{\nabla}^2(m) \right]\,\boldsymbol{K}_{\slashed{\nabla}^2(m)}(\boldsymbol{r}, \boldsymbol{r}', \tau) = 0 \quad \text{for}~\tau > 0, \end{equation} and supplemented by the initial condition
\begin{equation}\label{initial condition} 
\lim_{\tau \to 0} \boldsymbol{K}_{\slashed{\nabla}^2(m)}(\boldsymbol{r}, \boldsymbol{r}', \tau) = \delta(\boldsymbol{r},\boldsymbol{r}') \,\mathbb{1}. 
\end{equation}
Here, $\delta(\boldsymbol{r},\boldsymbol{r}')$ denotes the invariant delta function on $M^{D}$.  
Similarly to $\slashed{\nabla}^2(m)$, the spinor heat kernel $\boldsymbol{K}_{\slashed{\nabla}^2(m)}(\boldsymbol{r}, \boldsymbol{r}', \tau)$ is represented as a ${2^{[N/2]} \times 2^{[N/2]}}$ matrix. In curved or topologically nontrivial spacetimes, where eigenvalues are often not explicitly defined, the local heat kernel offers a more suitable approach, as it embodies essential information about the spacetime in which it is defined \cite{Hawking1977, Vassilevich2003}. In the subsequent section, we delve into a model of spacetime with nontrivial topology that illustrates the applicability of the local approach.

\section{Spinor fields on a cosmic dispiration spacetime} \label{S4}

Let us analyze a massive spinor field in the background of a four-dimensional Euclidean spacetime induced by the combined presence of a cosmic string and a screw dislocation, commonly referred to as cosmic dispiration spacetime \cite{Galtsov1993}. This background can be covered by cylindrical coordinates ${\{x^\mu\} = \{t, \rho, \phi, z\}}$, so that the associated line element takes the same form as Eq. \eqref{Generic Line Element}, expressed as
\begin{equation}\label{Line Element SC}
    ds^2 = dt^2 + d\rho^2 + \rho^2 d\phi^2 + (dz + \kappa d\phi)^2 .
    \end{equation} 
%Building on the formalism outlined in the previous section, we now examine a four-dimensional Euclidean spacetime whose line element takes the same form as Eq. \eqref{Generic Line Element}, expressed as:
%\begin{equation}\label{Line Element}
 %   ds^2 = dt^2 + d\rho^2 + \rho^2 d\phi^2 + dz^2 , 
 %   \end{equation}
Here, ${\rho\geq 0}$ and ${-\infty < z < \infty}$, while the angular coordinate $\phi$ takes values in the range ${0 \leq\phi \leq 2\pi/q}$. The parameter ${q}$ encodes the planar deficit angle, ${2\pi - 2\pi/q}$, and $\kappa$ denotes a constant associated with the screw dislocation, distinguishing this geometry from the usual Minkowski spacetime. %This conical spacetime arises from an infinitely long straight cosmic string aligned along the $z$-axis, with a linear mass density $\mu_0$. 
In the weak-field approximation, the relationship between $q$ and the linear mass density of the cosmic string $\mu_0$ is given by ${q^{-1} = 1 - 4\mu_0}$ \cite{Herondy2018}. %where $G$ denotes Newton’s gravitational constant \cite{Herondy2018}. 
Furthermore, spacetime is endowed with the topology $\mathbb{S}^1(\text{time}) \times M^3$, resulting from the compactification of the imaginary time coordinate $t$ into a circumference of length $\beta = T^{-1}$.

%We have presented a conceptually simple and efficient approach for calculating the spinor Casimir energy, as well as its corresponding temperature corrections, just by obtaining the spinor heat kernel associated with the operator $\slashed{\nabla}^2(m)$ defined on the base space $M^3$, whose nontrivial topology enforces a quasiperiodic boundary condition on the spinor field given by:

We now direct our attention to the base space $M^3$, where the massive operator $\slashed{\nabla}^2(m)$ is defined. It satisfies the following eigenvalue equation
\begin{align}\label{spatial spinor field}
   \slashed{\nabla}^2(m)\boldsymbol{\psi}_{a}(\boldsymbol{r}) = \left(\lambda_{a}^2 + m^2 \right)\boldsymbol{\psi}_{a}(\boldsymbol{r}) ,
\end{align}
with $\boldsymbol{\psi}_a(\boldsymbol{r})$ representing the spatial component of the spinor field, as defined through the time decomposition in Eq. \eqref{time decomposition}. 
%In particular, when ${q=1}$, the coordinate $\phi$ must exhibit a periodicity of $2\pi$, requiring the identification of $\phi$ and $\phi + 2\pi$. This results in an antiperiodic boundary condition for the spinor field, $\theta = 1/2$, 
%\begin{align}\label{Quasiperiodicity} \boldsymbol{\psi}_a(\rho, \phi, z) = -\boldsymbol{\psi}_a(\rho, \phi + 2\pi, z),
%\end{align}
%To enrich the discussion of the problem, we enforce a quasiperiodic boundary condition on the spinor field, expressed as:
We shall adopt a more generalized approach by imposing a quasi-antiperiodic boundary condition on the spinor field along the azimuthal direction, which can be formulated as
\begin{align}\label{Quasiperiodicity}
\boldsymbol{\psi}_a(\rho, \phi, z) = -e^{2\pi i\chi} \,\boldsymbol{\psi}_a(\rho, \phi + 2\pi/q, z), 
\end{align}  
where $\chi$ denotes a constant phase parameter constrained by the condition ${|\chi| \leq 1/2}$. In the absence of deficit angle (${q=1}$), the coordinate $\phi$ retains a periodicity of $ 2\pi$. Consequently, when the phase is set to ${\chi = 0}$, the field obeys an antiperiodic boundary condition due to the inherent sign inversion of spinor fields under a $ 2\pi$ rotation.

An interesting aspect of the geometry described by \eqref{Line Element SC} is that the spinor solutions $\boldsymbol{\psi}_a(\boldsymbol{r})$ to the above eigenvalue equation can be determined analytically on $M^3$, allowing for the explicit computation of the eigenvalues.
To proceed, we first define a frame. An appropriate choice of the orthonormal frame for this spacetime is as follows
\begin{align}\label{Frame field SD}
    \boldsymbol{X}_1 = \partial_{t}, \, \boldsymbol{X}_{2} = \partial_{\rho}, \, \boldsymbol{X}_3 = \frac{1}{\rho} (\partial_{\phi} -\kappa \partial_z), \, \boldsymbol{X}_4 = \partial_{z} .
\end{align}
In this frame, the metric components are ${\boldsymbol{g}(\boldsymbol{X}_\alpha, \boldsymbol{X}_\beta) = \delta_{\alpha\beta}}$,  ensuring that the line element matches the form presented in {Eq.~\eqref{Constant metric component}}. Moreover, by applying the spinor covariant derivative to this frame, it follows from \eqref{Spin connection} that the only nonvanishing spin connection components are
\begin{align}
    \omega_{323} = -\omega_{332} = \dfrac{1}{r} .
\end{align}
A convenient way to represent the frame \eqref{Frame field SD} is through the following Dirac matrices
%and $[N/2]$ being the floor of the number $N/2$. 
%\begin{align}\label{dirac.matrices.rep}
%\gamma_{1} &= \sigma_1\otimes \mathbb{1} , \quad \gamma_{2}  = \sigma_2\otimes \sigma_1 , \nonumber\\
%\gamma_{3} &= \sigma_2\otimes \sigma_2 , \quad \gamma_{4} = \sigma_2\otimes \sigma_3 ,
%\end{align} 
\begin{align}
\gamma_{i}  = \sigma_2\otimes \sigma_{i} , \quad \gamma_4 = \sigma_3\otimes \mathbb{1}\quad  \quad i=1,2,3, 
\end{align} 
%\begin{align}\label{dirac.matrices.rep}
%\gamma_{1} &= \sigma_1\otimes \mathbb{1} , \quad \gamma_{i}  = \sigma_2\otimes \sigma_{i-1} ,  \quad i=2,3,4, 
%\end{align} 
where 
\begin{eqnarray}
\sigma_1 = \left[
           \begin{array}{cc}
             0 & 1 \\
             1 & 0 \\
           \end{array}
         \right] ,   \sigma_2 =\left[
                                 \begin{array}{cc}
                                   0 & -i \\
                                   i & 0 \\
                                 \end{array}
                               \right] ,  \sigma_3 =  \left[
                                 \begin{array}{cc}
                                   1 & 0 \\
                                   0 & -1 \\
                                 \end{array}
                               \right],
\end{eqnarray}
are the hermitian Pauli matrices. These Dirac matrices can be easily verified to satisfy the Clifford algebra defined in Eq. \eqref{Clifford algebra}. 

Taking all these considerations into account, we now proceed to compute $\slashed{\nabla}^2(m)$, which is notably nontrivial in the context of curved space. In our case, after careful algebraic manipulation, we arrive at the following result    
\begin{align}\label{nabla^2(m) SD}
     \slashed{\nabla}^2(m) &= -\frac{1}{\rho}\left(\rho\,\partial_{\rho} \right) -\frac{1}{\rho^2}\,(\partial_{\phi} -\kappa \partial_z)^2 - \partial_z^2 \nonumber\\
   & + \frac{i}{\rho^2}\Sigma_3 (\partial_{\phi} -\kappa \partial_z) + \frac{1}{4\rho^2} + \frac{1}{4} R + m^2,
\end{align}
which depends on the diagonal matrix ${\Sigma_3 = \mathbb{1} \otimes \sigma_3}$ and the Ricci scalar $R$. For an idealized cosmic dispiration spacetime, 
the curvature vanishes everywhere except at ${\rho = 0}$, i.e., $\delta(\rho)$.
At this point, we require that the eigenfunctions remain regular for any $\rho>0$, that is outside the string.  The contribution of irregular eigenfunctions is discussed in Ref. \cite{Aram2013}.

Under the action of $\Sigma_3$, the field $\boldsymbol{\psi}_{a}(\boldsymbol{r})$ can be decomposed into two spinors corresponding to eigenvalues ${s_1 = \pm 1}$. Since the chiral matrix ${\gamma^5 = \sigma_1 \otimes \mathbb{1}}$ commutes with $\Sigma_3$ and $\slashed{\nabla}^2(m)$, these spinors are simultaneously eigenfunctions of $\gamma^5$, characterized by eigenvalues ${s_2 = \pm 1}$, representing the two possible chiralities. Together, these spinor indices define four independent solutions up to arbitrary 
constants.  
%which correspond precisely to the dimension of the spinor space in four dimensions.

%A convenient way to express these solutions is through the spinor basis $\{ \boldsymbol{\xi}^{+,s_1},  \boldsymbol{\xi}^{-,s_1}\}$, explicitly given by
A convenient way to represent these solutions, which respect both the spacetime symmetries and the condition \eqref{Quasiperiodicity}, is as follows
\begin{align}\label{Ansatz SD}
    \boldsymbol{\psi}_{a}(\boldsymbol{r}) &=   \mathcal{N} e^{i k\phi + i \nu z}\,\psi_{s_1, s_2}(\rho)\,\boldsymbol{\xi}_{s_1,s_2}, 
\end{align}
where 
\begin{align}
    k = q\left(\ell +1/2 + \chi \right), \quad \ell \in \mathbb{Z}. 
\end{align}
This stems from the fact that $\partial_z$ and $\partial_\phi$ are obvious Killing vector fields of our metric. Here, $\nu$ and ${\ell}$ correspond to continuous and discrete quantum numbers associated with linear and angular momentum, respectively, while $\mathcal{N}$ denotes the normalization constant. We also introduce the spinor basis $\{\boldsymbol{\xi}_{s_1,s_2}\}$, explicitly defined as
\begin{align}
    \boldsymbol{\xi}_{+,s_2} = \left[
                                 \begin{array}{c}
                                   1 \\
                                   0 \\
                                   s_2\\
                                   0\\
                                 \end{array}
                               \right], \quad \boldsymbol{\xi}_{-,s_2} = \left[
                                 \begin{array}{c}
                                   0 \\
                                   1 \\
                                   0\\
                                   s_2\\
                                 \end{array}
                               \right] ,
\end{align}
which are simultaneous eigenfunctions of $\Sigma_3$, with eigenvalue $s_1$, and $\gamma^5$ with eigenvalue $s_2$.
%Note that the spinors $\boldsymbol{\psi}_{a}(\boldsymbol{r})$ are eigenfunctions of the projection of the total angular 
%momentum ${J = - i\partial_{q\phi} + \Sigma_3/2}$ with eigenvalues determined by the equation:
%\begin{align}
%J \boldsymbol{\psi}_{a}(\boldsymbol{r}) = j \,\boldsymbol{\psi}_{a}(\boldsymbol{r}), \quad j = \ell+ \frac{s_1}{2} . 
%\end{align}

With this, by substituting the ansatz \eqref{Ansatz SD} into \eqref{spatial spinor field}, and using \eqref{nabla^2(m) SD}, we get 
the following ordinary differential equation for the radial components $\psi_{s_{1} s_{2}}(\rho)$
\begin{align}
    \left[\frac{d^2}{d\rho^2}  + \frac{1}{\rho} \frac{d}{d\rho} + \sigma^2 -\frac{\mu_{s_1}^2(\ell, \nu)}{\rho^2} \right]\psi^{s_{1} s_{2}}(\rho) 
= 0 , 
\end{align}
where the function $\mu_{s_1}(\ell, \nu)$ is defined as
\begin{align}\label{Bessel order}
    \mu_{s_1}(\ell, \nu) = q\left(\ell + 1/2+ \chi \right) -\kappa \nu - \frac{s_1}{2} ,
\end{align}
and the the parameter $\sigma $ satisfies the relation
 \begin{align}\label{Eigenvalues}
    \lambda_a^2 = \nu^2  + \sigma^2 . 
\end{align}
The regular solutions near the origin are expressed in terms of Bessel functions of the first kind,
${\psi_{s_1, s_2}(\rho) = J_{|\mu_{s_1}(\ell, \nu)|}(\sigma \rho)}$,
where the order $\mu_{s_1}(\ell, \nu)$ depends on the continuous quantum number $\nu$, as given by Eq.~\eqref{Bessel order}.
Although we neglect the singular point at $\rho = 0$, the solution must remain regular in its vicinity, leading us to discard any contributions from Neumann functions.
The label $a$ in the spinor field \eqref{Ansatz SD} then represents the collection of quantum numbers ${a = (\nu, \sigma, \ell, s_1, s_2)}$, which fully specify the spinor eigenfunctions.
This complete set of solutions satisfies the following orthonormality and completeness relations
\begin{align}
   \int d^{3}\boldsymbol{r}\,\sqrt{g^{(3)}}  \boldsymbol{\psi}_{a}(\boldsymbol{r})^{\dagger}  \boldsymbol{\psi}_{a'}(\boldsymbol{r}) = \delta_{aa'} , \label{Orthonormality} \\
   \sum_{a} \boldsymbol{\psi}_{a}(\boldsymbol{r})  \boldsymbol{\psi}_{a}(\boldsymbol{r}')^{\dagger} = \delta(\boldsymbol{r}, \boldsymbol{r}') \,\mathbb{1} ,\label{Completeness}
\end{align}
which uniquely determines the value of the normalization constant $\mathcal{N}$ in {Eq.~\eqref{Ansatz SD}}. The delta symbol in Eq.~\eqref{Orthonormality} should be interpreted as the Dirac delta function for continuous quantum numbers associated with the collective index $a$, and as the Kronecker delta for discrete ones. For the summation over $a$ in \eqref{Completeness}, we employ the compact notation defined below
\begin{align}\label{Summation}
    \sum_a = \int_{-\infty}^{\infty} d\nu  \int_{0}^{\infty} d\sigma \sum_{\ell=-\infty}^{\infty} \sum_{s_1= \pm}  \sum_{s_2= \pm} .
\end{align}
Hence, the complete set of normalized spinor solutions is given by
\begin{align}\label{Complete solution}
    \boldsymbol{\psi}_{a}(\boldsymbol{r}) =   \sqrt{\frac{q \sigma}{8 \pi^2}} e^{i(\ell + 1/2+\chi) q \phi + i k z}\,J_{|\mu_{s_1}(\ell, \nu)|}(\sigma \rho)\,\boldsymbol{\xi}_{s_1,s_2} .
\end{align}

With these solutions and the corresponding eigenvalues in \eqref{Eigenvalues}, we are now ready to derive an analytical expression for the local heat kernel associated with the massive operator $\slashed{\nabla}^{2}(m)$ in $M^3$, as defined by the expressions \eqref{heat equation} and \eqref{initial condition}.

\section{Spinor heat kernel and Casimir energy density} \label{S5}

In the scenario where the eigenfunctions and eigenvalues of $\slashed{\nabla}^{2}(m)$ are explicitly determined, the spinor heat kernel $\boldsymbol{K}_{\slashed{\nabla}^{2}(m)}$ admits the following local representation
\begin{align}\label{Local representation}
    \boldsymbol{K}_{\slashed{\nabla}^{2}(m)}(\boldsymbol{r}', \boldsymbol{r}, \tau) = e^{-\tau m^2} \sum_{a} e^{\lambda_{a}^2 \tau} \boldsymbol{\psi}_{a}(\boldsymbol{r})  \boldsymbol{\psi}_{a}^{\dagger}(\boldsymbol{r}') ,
\end{align}
where the summation over $a$ follows the definition provided in Eq. \eqref{Summation}.
It can be verified that the above expression not only solves Eq. \eqref{heat equation} but also satisfies the initial condition \eqref{initial condition}, as the spinor field obeys the completeness property outlined in Eq. \eqref{Completeness}. 

By substituting the eigenfunctions from {Eq.~\eqref{Complete solution}} along with the eigenvalues from {Eq.~\eqref{Eigenvalues}} into the spinor heat kernel expression in {Eq.~\eqref{Local representation}}, the following matrix structure is obtained
%\begin{widetext}
%\begin{align}
%   & K_{\slashed{\nabla}^{2}(m)}(\boldsymbol{r}', \boldsymbol{r}, \tau) = e^{-\tau m^2} \int_{-\infty}^{\infty} d\nu  \int_{0}^{\infty} d\sigma \,\sigma \sum_{\ell=-\infty}^{\infty}  e^{-\tau (\nu^2 +\sigma^2)} e^{i[\nu \Delta z + (\ell + \theta)q\Delta \phi]}\times \nonumber\\
%   &\left[
 %                                \begin{array}{cccc}
  %                                 J_{|\mu_{+}(\ell, \nu)|}(\sigma \rho) J_{|\mu_{+}(\ell, \nu)|}(\sigma \rho')& 0& 0& 0 \\
 %                                  0& J_{|\mu_{-}(\ell, \nu)|}(\sigma \rho) J_{|\mu_{-}(\ell, \nu)|}(\sigma \rho')& 0& 0 \\
   %                                0& 0& J_{|\mu_{+}(\ell, \nu)|}(\sigma \rho) J_{|\mu_{+}(\ell, \nu)|}(\sigma \rho')& 0 \\
   %                                0& 0& 0& J_{|\mu_{-}(\ell, \nu)|}(\sigma \rho) J_{|\mu_{-}(\ell, \nu)|}(\sigma \rho') \\
    %                             \end{array}
     %                          \right] ,
%\end{align}
%\end{widetext}
%\begin{align}\label{Matrix heat kernel}
 %   K_{\slashed{\nabla}^{2}(m)}(\boldsymbol{r}', \boldsymbol{r}, \tau)  = \left[
  %                               \begin{array}{cccc}
  %                                 K_{+}& 0& 0& 0 \\
  %                                 0&  K_{-}& 0& 0 \\
  %                                 0& 0&  K_{+}& 0 \\
  %                                 0& 0& 0&  K_{-} \\
  %                               \end{array}
  %                             \right] ,
%\end{align}
\begin{align}\label{Matrix heat kernel}
    \boldsymbol{K}_{\slashed{\nabla}^{2}(m)}(\boldsymbol{r}', \boldsymbol{r}, \tau)  = \text{diag}(K_{+}, K_{-}, K_{+}, K_{-}),
\end{align}
where the four nontrivial components of the spinor heat kernel, ${K_s = K_s(\boldsymbol{r}', \boldsymbol{r}, \tau)}$, are expressed in the form
%\begin{widetext}
\begin{align}
  & K_{s} = \frac{q\,e^{-\tau m^2}}{4\pi^2} \int_{-\infty}^{\infty} d\nu\,e^{-\tau \nu^2 + i\nu \Delta z} \sum_{\ell=-\infty}^{\infty} e^{iq (\ell + 1/2+\chi)\Delta \phi} \nonumber\\
  &\times   \int_{0}^{\infty} d\sigma \sigma e^{-\tau \sigma^2}  J_{|\mu_{s}(\ell, \nu)|}(\sigma \rho) J_{|\mu_{s}(\ell, \nu)|}(\sigma \rho') .
  \end{align}
%\end{widetext}
Here, ${\Delta z = z' - z}$, ${\Delta \phi = \phi' -\phi}$, and the index $s$ assumes the values ``$\pm$''. To perform the integration on $\sigma$, which involves a product of Bessel functions, we employ the following identity \cite{Prudnikov1986}
\begin{align}
    \int_{0}^{\infty}d\sigma  \sigma e^{-\tau \sigma^2} J_{|\mu_{s}(\ell, \nu)|}(\sigma \rho) J_{|\mu_{s}(\ell, \nu)|}(\sigma \rho') \nonumber\\
    = \frac{1}{2\tau} e^{- \frac{\rho^2 + \rho'^2}{4\tau} } I_{|\mu_{s}(\ell, \nu)|}\left(\frac{\rho \rho'}{2\tau}\right) ,
\end{align}
where $I_{\mu}(x)$ denotes the modified Bessel function of order $\mu$. As a result, the expression for $K_{s}$ simplifies to the following form
\begin{align}
  K_s & = \frac{q}{8\pi^2 \tau}e^{-\frac{\rho^2 + \rho'^2}{4\tau}-\tau m^2}\int_{-\infty}^{\infty}d\nu e^{-\tau \nu^2 + i\nu \Delta Z} \times \nonumber\\
  &\times e^{\frac{is \Delta \phi}{2}} \sum_{\ell=-\infty}^{\infty} e^{i\mu_{s}(\ell, \nu)\Delta \phi}I_{|\mu_{s}(\ell, \nu)|}\left(\frac{\rho \rho'}{2\tau}\right) ,
\end{align}
where $\Delta Z = \Delta z + \kappa \Delta \phi$. 
%Furthermore, performing the substitution $r = \frac{\rho \rho'}{2\tau}$, we can rewrite it in a very compact form:
Furthermore, by rewriting it in the compact form
\begin{align}\label{Appendix 1}
  K_s  = \frac{q}{8\pi^2 \tau}e^{-\frac{\rho^2 + \rho'^2}{4\tau}-\tau m^2}\,\mathcal{I}_s(r, q, \kappa), 
\end{align}
we can see that explicitly determining the spinor heat kernel requires evaluating $\mathcal{I}_s(r, q, \kappa)$, which involves the integral over the continuous quantum number $\nu$ in the following form
\begin{align}\label{Integral}
    \mathcal{I}_s(r, q, \kappa) = \int_{-\infty}^{\infty}d\nu\,e^{-\tau \nu^2 + i\nu \Delta Z} \mathcal{S}_s(r, q, \kappa) , 
\end{align}
and $ \mathcal{S}_s(r, q, \kappa)$, which corresponds to the summation over the discrete quantum numbers $\ell$ given below
\begin{align}\label{Summation2}
    \mathcal{S}_s(r, q, \kappa) =  e^{\frac{is \Delta \phi}{2}} \sum_{\ell=-\infty}^{\infty} e^{i\mu_{s}(\ell, \nu)\Delta \phi}I_{|\mu_{s}(\ell, \nu)|}(r) , 
\end{align}
where ${r = \frac{\rho \rho'}{2\tau}}$. 
When the order of the Bessel function depends on the continuous quantum number $\nu$, the evaluation of the integral in \eqref{Integral} and the summation in \eqref{Summation2} becomes considerably more challenging, as discussed in Appendix \ref{Heat kernel}.

Before addressing the general case, it is instructive to first examine the simpler situation with ${\kappa = 0}$, corresponding to the pure cosmic string.
%For this case, the thermal effects on the spinor vacuum energy density have recently been explored in the energy-momentum tensor context in Ref. \cite{Bezerra2024}. However, within the framework of the heat-kernel approach to zeta-function regularization, these thermal effects associated with spinor fields under quasiperiodic boundary conditions have not yet been discussed in the literature. Therefore, the following section not only revisits the pure cosmic string case using this robust and elegant method but also introduces a framework that provides a more effective understanding of the screw dislocation's effects on the heat kernel coefficients and, consequently, on the vacuum-free energy.
In this case, although thermal effects on the spinor vacuum energy density have been investigated in \cite{Bezerra2024}, they have not yet been explored under quasi-antiperiodic boundary conditions within the heat kernel framework. The following section not only revisits this setup but also develops a framework for analyzing the influence of the screw dislocation on the heat kernel coefficients and the vacuum free energy.

\subsection{Case $\kappa = 0$}\label{kappa zero}

In the limiting case $\kappa = 0$, the background $\eqref{Line Element SC}$ simplifies to the standard scenario of an idealized cosmic string along the $z$-axis, with the corresponding line element given by
\begin{align}\label{Line element CS}
    ds^2 = dt^2 +d\rho^2 + \rho^2 d\phi^2 + dz^2 . 
 \end{align}
In addition, the Bessel function order \eqref{Bessel order} no longer depends on the continuous quantum number $\nu$. This allows the function $\mathcal{S}_s(r, q, 0)$ to be factored out of the integral in \eqref{Integral}, thereby reducing it to a Gaussian-type integral that can be easily computed. 
Also, the evaluation of the remaining summation in $\mathcal{S}_s(r, q, 0)$ is a well-established procedure, as can be consulted in {Refs.~\cite{Aram2015, Azadeh2021, Belluci2014, Valdir2006}}. In particular, Appendix A from {Ref.~\cite{Aram2015}} provides a convenient integral representation of the function $\mathcal{S}_s(r, q, 0)$, which results in the following expression for $K_{s}$
%The procedure used to compute each of the heat kernel components outlined above has been extensively examined in numerous studies, such as those cited in Refs. \cite{Aram2015, Azadeh2021, Herondy2024, Valdir2006}. Utilizing the techniques proposed in these works, we derived the following results:
%\begin{widetext}
%\begin{align}\label{K_AA CS 1}
 % K_{AA}^{s} & = \frac{e^{-\tau m^2}}{(4\pi \tau)^{3/2}} \left[\sum_{\ell} \frac{e^{2\pi i \ell \left(\theta+ \frac{s}{2q}\right) - \frac{i s \Delta \phi}{2}}}{e^{\frac{\Delta R_{\ell}^2}{4\tau}}} - \frac{q}{2\pi i}\sum_{b=\pm 1} b e^{i b q \pi \ell \left(\theta+ \frac{s}{2q}\right) + i q \theta \Delta \phi}\right.\times \nonumber\\
 % &\times\left.\int_{0}^{\infty} dy \,\frac{\cos\left[qy\left(1-\theta - \frac{s}{2q}\right) \right]-\cos\left[qy\left(\theta + \frac{s}{2q}\right) \right]e^{-iq(\Delta \phi + sb\pi)}}{e^{\frac{\Delta R_{y}^2}{4\tau}} [\cosh(qy) - \cos(q\Delta \phi + qb\pi)] }\right] . 
%\end{align}
% \end{widetext}
\begin{align}\label{K_AA CS 1}
&  K_{s} = \frac{e^{-\tau m^2 - \frac{i s \Delta \phi}{2}}}{(4\pi \tau)^{3/2}} \left[\sum_{\ell} \frac{e^{2\pi i \ell \chi_s}}{e^{\frac{\Delta R_{\ell}^2}{4\tau}}} -\sum_{b=\pm 1}\frac{bqe^{i q \chi_s (\Delta \phi+ b \pi )}}{2\pi i} \right.\nonumber\\
  &\times\left.\int_{0}^{\infty} dy \,\frac{\cos\left[qy\left(1-\chi_s\right) \right]-\cos(qy \chi_s)e^{-iq(\Delta \phi + b\pi)}}{e^{\frac{\Delta R_{y}^2}{4\tau}} [\cosh(qy) - \cos(q\Delta \phi + qb\pi)] }\right] . 
\end{align}
The summation over $\ell$ is carried out under the condition that 
\begin{align}\label{Summation over l}
    -\frac{q}{2} + \frac{q\Delta \phi}{2\pi} \leq \ell \leq \frac{q}{2} + \frac{q\Delta \phi}{2\pi} ,
\end{align}
and the quantities $\Delta R_{\ell}^2, \Delta R_{y}^2$ and $\chi_s$ are defined as
\begin{align}
    \Delta R_{\ell}^2 &= \Delta z^2 + \rho^2 + \rho'^2 -2\rho\rho' \cos\left(\frac{2\pi \ell}{q} -\Delta\phi\right),\nonumber\\
    \Delta R_{y}^2 &= \Delta z^2 + \rho^2 + \rho'^2 +2\rho\rho' \cosh(y), \nonumber\\
    \chi_s &= \chi +\frac{1}{2} +\frac{s}{2q} .
\end{align}

Note that for ${1 \leq q< 2}$, the only contribution in summation over $\ell$ in Eq. \eqref{K_AA CS 1} comes from the term ${\ell = 0}$ at the coincidence limit ${\boldsymbol{r}' \rightarrow \boldsymbol{r}}$. This term corresponds to the Euclidean contribution to the spinor heat kernel function.
Thus, upon taking the coincidence limit and executing the summation over $b$, it is convenient to express $K_{s}$ as
\begin{align}\label{Notation [q/2]}
  K_{s} & = \frac{e^{-\tau m^2}}{(4\pi \tau)^{3/2}} \left[1 + 2\sum_{\ell=1}^{[q/2]}\cos\left(2\pi \ell \chi_s\right) e^{-\frac{(\rho u_{\ell})^2}{\tau}} \right.\nonumber\\ 
  &-\left.\frac{q}{\pi}\int_{0}^{\infty}\frac{f_s(q, y, \chi)}{ \cosh(qy) - \cos(q\pi) }\,e^{-\frac{(\rho u_{y})^2}{\tau}} \right] . 
\end{align}
Here, the function $f_{s}(q, y, \chi)$ is defined as
 \begin{align}
 f_s(q, y, \chi) &=  \cosh\left[qy\left(1-\chi_s\right) \right]\sin\left(q\pi\chi_s \right)\nonumber\\
 &+ \cosh\left(qy \chi_s \right)\sin\left[q\pi\left(1-\chi_s\right) \right], 
 \end{align}
and quantities $u_\ell$ and $u_y$ are given by
\begin{align}
    u_{\ell} = \sin\left(\pi \ell/q\right), \quad u_y = \cosh(y/2). 
\end{align}
The notation $[q/2]$ in {Eq.~\eqref{Notation [q/2]}} denotes the floor of the number $q/2$. In particular, when $q$ is an even integer, the associated terms in the summation must be adjusted with an additional factor of $1/2$. 
Therefore, by inserting the components $K_s$ into {Eq.~\eqref{Matrix heat kernel}}, the spinor heat kernel can be decomposed into two distinct terms, as follows
 \begin{align}\label{Local HQ CS}
    \boldsymbol{K}_{\slashed{\nabla}^{2}(m)}(\boldsymbol{r}, \boldsymbol{r}, \tau) &= \boldsymbol{K}^{\mathbb{E}}(\boldsymbol{r}, \boldsymbol{r}, \tau) \nonumber\\
     &+ \text{diag}(K^{\mathbb{T}}_{+}, K^{\mathbb{T}}_{-}, K^{\mathbb{T}}_{+}, K^{\mathbb{T}}_{+}).
 \end{align}
The first term, $\boldsymbol{K}^{\mathbb{E}}(\boldsymbol{r}, \boldsymbol{r}, \tau)$, corresponding to ${\ell=0}$ in the series, represents the Euclidean local heat kernel contribution given by
\begin{align}\label{Euclidean C}
    \boldsymbol{K}^{\mathbb{E}}(\boldsymbol{r}, \boldsymbol{r}, \tau) =   \frac{e^{-\tau m^2}}{(4\pi \tau)^{3/2}} \mathbb{1}.
\end{align}
The second term corresponds to the topological  contribution along with the quasi-antiperiodic condition, whose components $ K^{\mathbb{T}}_{s}(\boldsymbol{r}, \boldsymbol{r}, \tau)$ are defined as
%\begin{align}
 %    K^{\text{T}}_{s}(\boldsymbol{r}, \boldsymbol{r}, \tau) = K_{s}(\boldsymbol{r}, \boldsymbol{r}, \tau) - K^{E}(\boldsymbol{r}, \boldsymbol{r}, \tau).  
%\end{align}
\begin{align}
     K^{\mathbb{T}}_{s} &= \frac{e^{-\tau m^2}}{(4\pi \tau)^{3/2}} \left[2\sum_{\ell=1}^{[q/2]}\cos\left(2\pi \ell \chi_s\right) e^{-\frac{(2 \rho u_{\ell})^2}{4\tau}} \right.\nonumber\\ 
  &-\left.\frac{q}{\pi}\int_{0}^{\infty}dy\frac{f_{s}(q, y, \chi)}{ \cosh(qy) - \cos(q\pi) }\,e^{-\frac{(2\rho u_{y})^2}{4\tau}} \right] . 
\end{align}

In particular, this convenient subdivision into Euclidean and topological contributions is reflected in the structure of the asymptotic expansion of the global heat kernel, defined in {Eq.~\eqref{heat equation expansion}}. According to {Eq.~\eqref{Global heat kernel}}, the global heat kernel $ K_{\slashed{\nabla}^2(m)}(\tau) $ is obtained by integrating ${\text{tr}\left[\boldsymbol{K}_{\slashed{\nabla}^2(m)}(\boldsymbol{r}, \boldsymbol{r}', \tau)\right]}$ over the whole $3$-dimensional space.
This procedure directly yields the following result
\begin{align}\label{Asymptotic expansion}
    K_{\slashed{\nabla}^2(m)}(\tau) = \frac{e^{-\tau m^2}}{(4\pi \tau)^{3/2}}(c_0 + c_1\tau),
\end{align}
where the spinor heat kernel coefficients, $c_0$ and $c_1$, for a quasi-antiperiodically identified cosmic string spacetime (CS)
are given by
\begin{align}
    c_0 &= 4 V_3 , \label{c_0} \\
    c_1(q, \chi) &= R_{\text{CS}}(q,\chi) V_1\nonumber\\
    &=\frac{8\pi}{q}\left[\sum_{\ell=1}^{[q/2]}(-1)^{\ell}\,\frac{\cos\left(2\pi \ell \chi\right)\cos\left(\ell \pi/q\right)}{u_{\ell}^2} \right.\nonumber\\ 
  &-\left.\frac{q}{4\pi}\int_{0}^{\infty}\frac{dy}{u_{y}^2} \frac{f_{+}(q, y, \chi)+f_{-}(q, y, \chi) }{\cosh(qy) - \cos(q\pi)}\right]  V_1 .  \label{c_1}
\end{align}
Here, $V_3$ represents the infinite volume of the $3$-dimensional base space $M^3$ and $V_1$ denotes an infinite length along the $z$-direction. 
The coefficient $c_0$ corresponds to the Euclidean heat kernel, while $c_1$ arises from both the nontrivial topology and the quasi-antiperiodic condition, as evidenced by its dependence on the deficit angle parameter $q$ and the phase parameter $\chi$. 

\begin{figure}
    \centering
    \hspace{-7.2cm}$(a)$\\
    \includegraphics[width=0.47\textwidth]{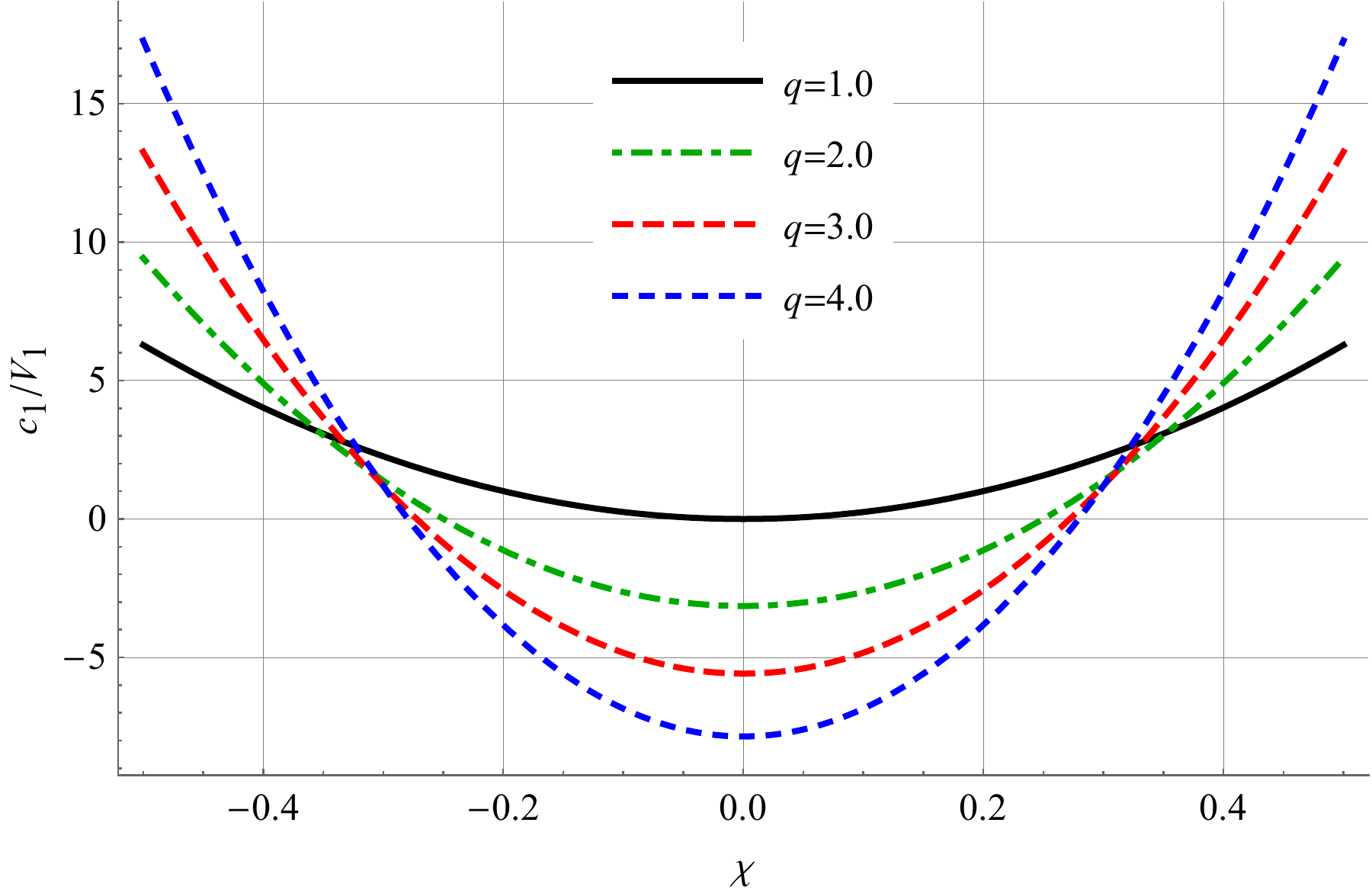}\\
     \hspace{-7.2cm}$(b)$\\
    \includegraphics[width=0.47\textwidth]{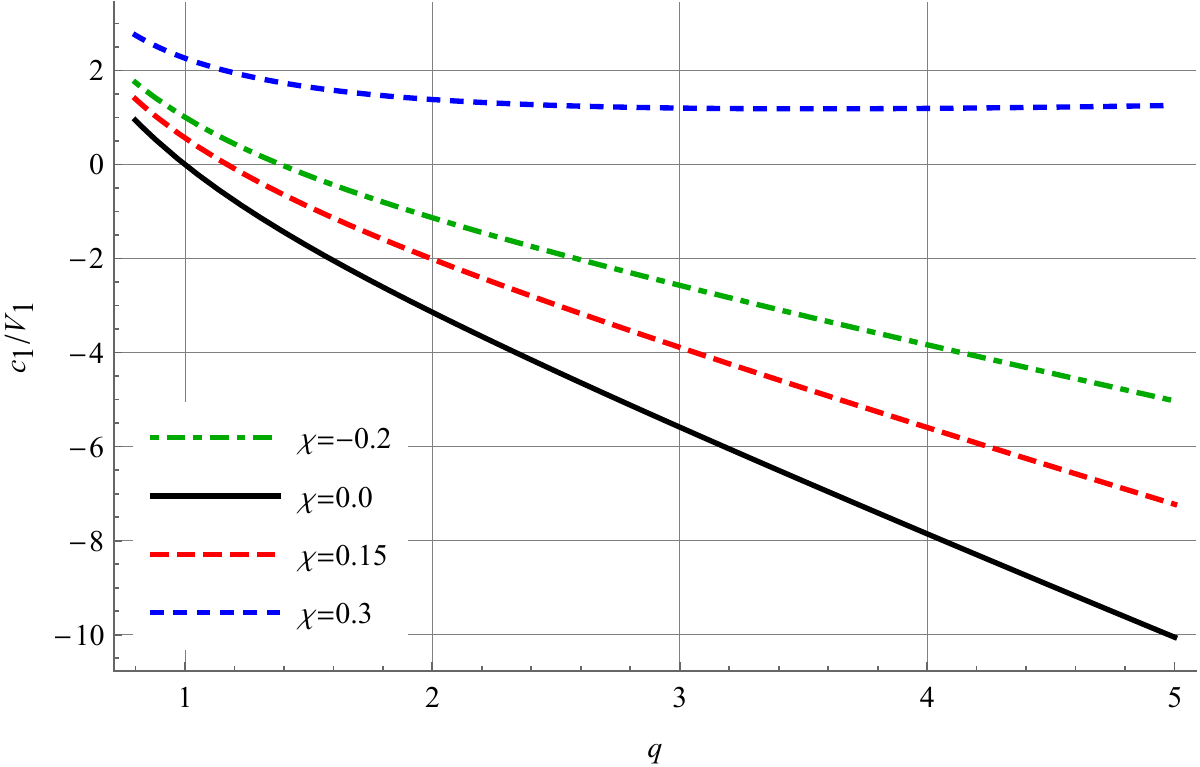}
    \caption{In $(a)$ the ratio ${R = c_1/V_1}$ is shown as a function of $\chi$ for various parameter values of $q$, and in $(b)$ it is represented as a function of $q$ for different values of $\chi$. In both cases, the heat kernel coefficient $c_1$ vanishes at ${q = 1}$ and ${\chi = 0}$.}
    %Ratio ${c_1/V_1}$ plotted as a function of $\theta$ for multiple parameter values $q$. The continuous curve corresponding to ${q = 1}$ goes to zero at  ${\theta = 1/2}$.} 
    \label{Plot c1 1}
\end{figure}

%In Fig. \ref{Plot c1 1}, the ratio $c_1/V_1$ is depicted as a function of the phase parameter $\theta$, with variations corresponding to different angle deficit parameter $q$ values. 
{Figure~\ref{Plot c1 1}}$(a)$ illustrates the ratio ${c_1/V_1}$ as a function of the phase parameter $\chi$, while {Fig.~\ref{Plot c1 1}}$(b)$ presents it as a function of the deficit angle parameter $q$. In both cases, the plots highlight variations that correspond to different values of these parameters.
For any value of $q$, the coefficient $c_1$ attains an absolute minimum at ${\chi = 0}$, thereby reducing the quasi-antiperiodic boundary condition given in {Eq.~\eqref{Quasiperiodicity}} to an antiperiodic one. This minimum value of $c_1$ vanishes when ${q = 1}$, i.e., with no deficit angle, and becomes increasingly negative as the parameter ${q > 1}$ grows larger, as illustrated in the figure.  
This result is consistent with expectations, as the geometry with ${q=1}$ and ${\chi=0}$ simplifies to the standard Minkowski spacetime, where the coordinate $\phi$ exhibits a periodicity of $2\pi$. The antiperiodic boundary condition for the spinor field arises naturally from the intrinsic sign inversion of spinor fields under a $2\pi$ rotation. 
Another remarkable feature of the plots is the presence of specific values of ${q>1}$ and $\chi$ for which the coefficient $c_1$ vanishes. This emphasizes that the quasi-antiperiodic condition effects can completely cancel out the topological contribution, leaving only the coefficient $c_0$ related to the Euclidean divergence.
%Due to this complex dependence of $c_1$ on the parameters $q$ and $\chi$, it is helpful to examine their contributions individually. 

%The proportional factor that directly connects the spinor coefficient $c_0$ to its corresponding scalar field is straightforward: it corresponds to the dimension of the spinor space \cite{Joas2024, Herondy2024}. In contrast, the complex dependence of $c_1$ on the parameters $q$ and $\chi$  makes determining its proportional factor more challenging. To establish this relation, it is helpful to examine the contributions associated with $q$ and $\chi$ individually. 
Let us provide some observations comparing the spinor heat kernel coefficients $c_0$ and $c_1$ with the scalar ones presented in \cite{Herondy2024}, expected to differ because of the spinor degrees of freedom. The coefficient $c_0$ is directly related to the scalar coefficient through a constant multiplicative factor, corresponding to the dimension of the spinor space \cite{Herondy2024, Joas2024}.
In contrast, the coefficient $c_1$ exhibits a more intricate dependence on the parameters $q$ and $\chi$, making it more difficult to determine a simple proportionality factor. To establish this relation, it is helpful to examine the contributions associated with $q$ and $\chi$ individually. 
We start by considering the contribution purely associated with the deficit angle parameter $q$, achieved by setting ${\chi = 0}$. In this case, the following relation holds
\begin{align}
   & f_{+}(q, y, 0)+f_{-}(q, y, 0) = \nonumber\\
    &- 4\cos\left(\frac{\pi q}{2}\right)\,\sinh\left(\frac{y}{2}\right) \sinh\left(\frac{qy}{2}\right) ,
\end{align}
which vanishes when $q$ takes odd values.
Furthermore, in the case where ${\chi =0}$ and $q$ is odd, it follows from {Eq.~\eqref{c_1}} that the summation over $\ell$ in the coefficient $c_1$ can be rewritten as
\begin{align}
    \sum_{\ell=1}^{[q/2]} \rightarrow \frac{1}{2}\sum_{\ell=1}^{q-1} . 
\end{align}
Under these conditions, the coefficient $c_1$  simplifies to:
\begin{align}\label{c1(q)}
    c_1(q, 0) &= \frac{4\pi}{q} \sum_{\ell=1}^{q-1} (-1)^\ell \,\frac{\cot\left(\frac{\pi\ell}{q}\right)}{\sin\left(\frac{\pi \ell}{q}\right)} \nonumber\\
    & = - 2 \left[ V_1\frac{\pi}{3} \left(q -\frac{1}{q}\right) \right],
\end{align}
which vanishes at ${q = 1}$, as it must for consistency.
A notable feature of the above expression is its analyticity in $q$, which enables it to be naturally extended to all values of $q$ through analytic continuation. This expression precisely matches the curve corresponding to ${\chi = 0}$ shown in Fig.~\ref{Plot c1 1}($b$). Moreover, it equals minus twice the corresponding coefficient associated with the scalar field presented in {Ref.~\cite{Herondy2024}}. In general, the ratio of the spinor heat coefficients $c_1$ to the corresponding scalar one is given by $``-2^{[N/2]-1}"$ on manifolds with conical singularities \cite{Fursaev1997, Fursaev1994}, and thus our results are consistent with the expected value in the four-dimensional case.

Now, the pure contribution arising from the phase parameter $\chi$ can be determined by setting ${q = 1}$. For this case, the summation over $\ell$ in the coefficient $c_1$ from {Eq.~\eqref{c_1}} vanishes. Consequently, {Eq.~\eqref{c_0}} simplifies to
\begin{align}
    &c_1(1, \chi) \nonumber\\
    &= -8 V_1 \int_{0}^{\infty} dy\,\frac{\cos\left[\pi (\chi + 1/2)\right] \sinh (y \chi) \tanh(y/2)}{1+\cosh(y)} \nonumber\\
        &= 8 \pi V_1 \chi^2 ,
\end{align}
which appropriately vanishes at ${\chi = 0}$, as expected.
This expression corresponds exactly to the curve for ${q=1}$ illustrated in {Fig.~\ref{Plot c1 1}}($a$). In this regard, incorporating the quasi-antiperiodicity effects extends the results presented in {Ref.~\cite{Fursaev1997}}  for spin-$1/2$ fields, resulting in a more comprehensive and robust analytical structure for coefficient $c_1$. In addition, it serves as a spinorial extension of the findings reported in {Ref.~\cite{Herondy2024}}.

The heat kernel coefficients $c_0$ and $c_1$ contribute to the divergent terms resulting from the geometry and topology of spacetime, which directly influence the generalized zeta function defined in Eq.~\eqref{integral representation}. Indeed, from {Eq.~\eqref{Asymptotic expansion}}, this function can be expressed as
\begin{align}\label{Gamma times zeta}
  &\,\Gamma\left(w- 1/2\right) \zeta_{\slashed{\nabla}^2(m)}\left(w- 1/2\right)\nonumber\\
      & =  \frac{1}{(4\pi)^{3/2}}\sum_{p =0}^{1} \frac{c_p\,\Gamma\left(w + p - 2 \right)}{ \left(m^2\right)^{w + p - 2}} , 
\end{align}
which has simple poles at $w= 0$, with residue 
\begin{align}\label{Residue}
     \text{Res}(\zeta_{\slashed{\nabla}^2(m)}\left(w - 1/2\right), w = 0) =  -\frac{c_2(m)}{16\pi^2} .
\end{align}
Here, the coefficient $c_2(m)$ follows the definition provided in {Eq.~\eqref{c_2(m) heat kernel coefficient}}, with ${c_2 = 0}$ imposed. Explicitly, this means
\begin{align}
    c_2(m) = \frac{m^4 c_0}{2} - m^2 c_1. 
\end{align}

From {Eq.~\eqref{Residue}}, it is evident that for nonvanishing heat kernel coefficients, the generalized zeta function is not finite, necessitating a renormalization procedure to address the divergence.
The Casimir energy, while finite due to the FP prescription in its definition, {Eq.~\eqref{Renormalized Casimir energy}}, includes terms involving coefficients $c_0$ and $c_1$ that increase without bounds with mass $m$. Contributions associated with $c_0$ and $c_1$ must be completely removed to ensure consistent renormalization and define unambiguous spinor vacuum free  energy. As a result, the renormalized Casimir energy associated with a massive spinor field defined on quasi-antiperiodically identified cosmic string spacetime reduces to zero. Explicitly
\begin{align}\label{RCE C1}
    E_{\text{C}}^{\text{ren}}(m) = 0.
\end{align}
This result trivially satisfies the condition outlined in Eq.~\eqref{renormalization condition}. In the massless case (${m = 0}$), as ${c_2}$ is naturally zero, $\zeta_{\slashed{\nabla}^2(m)}\left(-1/2\right)$ is finite, leading to complete vanishing of the Casimir energy. 
%As a result, the need for a renormalization procedure is entirely eliminated in this scenario. 
Consequently, the necessity for the renormalization condition \eqref{renormalization condition} is completely eliminated in this scenario. 
Also, by employing the local version of Eq. \eqref{integral representation} along with the local heat kernel from Eq. \eqref{Local HQ CS}, one can verify that the resulting renormalized Casimir energy density coincides with the expression provided in Ref. \cite{Belluci2014}.

\subsubsection{Temperature corrections}\label{Zeta}

Let us examine the temperature correction, $\Delta F(\beta, m)$, to the Casimir energy. 
Substituting the global heat kernel \eqref{Asymptotic expansion} into $\Delta F(\beta, m)$, as defined in {Eq.~\eqref{Generic temperature correction}}, allows the derivation of the Casimir thermal correction structure for the massive spinor field, given by
\begin{align}
    \Delta F(\beta, m)  =  \Delta F^{\mathbb{E}}(\beta, m) + \Delta F^{\mathbb{T}}(\beta, m) .
\end{align}
The term $\Delta F^{\mathbb{E}}(\beta, m)$ represents the contribution arising from the Euclidean heat kernel. It is associated with the coefficient $c_0$ and therefore remains independent of the topology parameter. Its explicit form is given by
\begin{align}\label{Euclidean TC}
    \Delta F^{\mathbb{E}}(\beta, m) = \frac{c_0}{2\pi^2\beta^4} \sum_{n=1}^{\infty} \frac{\cos(\pi n)}{n^2} (m\beta)^2 K_2(m\beta n) ,
\end{align}
where $K_{\mu}(z)$ is the MacDonald function \cite{Abramowitz1972}. The term $\Delta F^{\mathbb{T}}(\beta, m)$ is the contribution from topology and the quasi-antiperiodic condition, related to the coefficient $c_1$. It has the following form
\begin{align}\label{Topological TC}
    \Delta F^{\mathbb{T}}(\beta, m) = \frac{c_1}{4\pi^2\beta^2} \sum_{n=1}^{\infty} \frac{\cos(\pi n)}{n} (m\beta) K_1(m\beta n) .
\end{align}

In particular, the massless case can be obtained by making use of the following limit
\begin{align}\label{Small arguments limit}
    \underset{x \rightarrow 0}{\text{lim}}\,x^\mu  K_{\mu}(n x) = \left(\frac{2}{n}\right)^\mu  \frac{\Gamma(\mu)}{2} . 
\end{align}
In fact, by splitting the sum in {Eqs.~\eqref{Euclidean TC}} and \eqref{Topological TC} into even and odd terms, and applying the preceding equation, one can promptly verify the following massless limit
\begin{align}
    \underset{m \rightarrow 0}{\text{lim}}\,\Delta F^{\mathbb{E}}(\beta, m) &=  \frac{c_0}{4\pi^2\beta^4} \left[ \zeta(4) - \zeta\left(4, \frac{1}{2}\right)\right],\nonumber\\
    \underset{m \rightarrow 0}{\text{lim}}\,\Delta F^{\mathbb{T}}(\beta, m) &=\frac{c_1}{\pi^2\beta^2} \left[ \zeta(2) - \zeta\left(2, \frac{1}{2}\right)\right],
\end{align}
where $\zeta(x)$ is the standard Riemann zeta function and $\zeta(x, y)$ is the Hurwitz zeta function defined for $\text{Re}(x)> 1$ and $y \neq 0, -1, -2, \ldots$, in the form \cite{Elizalde1994}
\begin{align}\label{Hurwitz zeta}
    \zeta(x, y) = \sum_{n=0}^{\infty} (n + y)^{-x} .
\end{align}
%Note that Riemann zeta function is $\zeta(z, 1)$. 
By employing the relation
\begin{align}\label{HZF relation}
    \zeta\left(x, \frac{1}{2}\right) = (2^x -1) \,\zeta(x), 
\end{align}
along with the known values ${\zeta(4) = \pi^4/90}$ and ${\zeta(2) = \pi^2/6}$, it can be demonstrated that the Casimir thermal correction for a massless spinor field in the presented geometry conforms to the following structure
\begin{align}\label{Massless TC}
   \Delta F(\beta) =  \sum_{k= 0}^{2} \alpha_{k} T^{4-k},
\end{align}
where the $\alpha_k$ coefficients are expressed in
terms of the heat kernel coefficients as follows
\begin{align}\label{Alpha coefficients}
    \alpha_0 = \frac{7}{2}\left(-\frac{\pi^2 c_0}{90}\right) , \quad \alpha_1 = 0, \quad \alpha_2 = 2\left(-\frac{c_1}{24}\right) .
\end{align}
Moreover, their values are influenced by the field's spin, as discussed in {Ref.~\cite{Klimchitskaya2009}}. Specifically, the term ${k=0}$, associated with spinor black body radiation, has a coefficient $\alpha_0$ that is $7/2$ times greater than the corresponding scalar field coefficient \cite{Herondy2024, Joas2024}. On the other hand, the term ${k=2}$, arising from the effects of nontrivial topology and the quasi-antiperiodic condition, has a coefficient $\alpha_2$ that is twice the value found in {Ref.~\cite{Herondy2024}} for the scalar case.
%In fact, the term ${k=0}$ in the summation accounts for the contribution from spinor black body radiation. The associated coefficient, $\alpha_0$, for the spinor field is $7/2$ times higher than the corresponding coefficient for the scalar field \cite{Herondy2024, Joas2024}. Additionally, the term ${k=2}$, the only nonvanishing term in the summation, arises due to the influence of the nontrivial topology and the quasiperiodic condition. The corresponding coefficient, $\alpha_2$, is twice the value obtained in {Ref.~\cite{Herondy2024}} for the massless scalar case. 

Nonzero heat kernel coefficients indicate the presence of divergences. Even when finite, they introduce ambiguity in the zeta function regularization prescription due to its dependence on the arbitrary parameter $\mu$. Consequently, terms involving any heat kernel coefficient must be subtracted. Therefore, the renormalized temperature correction contribution reduces to zero,
explicitly
\begin{align}\label{RTE C1}
    \Delta F^{\text{ren}}(\beta, m) = 0 . 
\end{align}
In the massless scenario, renormalization is crucial to ensure a proper classical contribution in the high-temperature regime \cite{Joas2024}. As the Casimir effect is inherently a quantum phenomenon, power-law contributions in $T$, like \eqref{Massless TC}, are not expected to have a significant influence at high-temperature limit.

Building upon all these considerations, we are now ready to address the case where ${\kappa \neq 0}$, the cosmic dispiration spacetime. As will become clear shortly, the particularities associated with the zeta function method explored in detail in this section pave the path for a more effective understanding of this scenario.

\subsection{Case $\kappa \neq 0$}

We now examine the case of cosmic dispiration spacetime. Here, our focus is on the purely topological contribution, which is associated with the deficit angle $q$ and the screw dislocation parameter $\kappa$. This is achieved by setting ${\chi = 0}$ in Eq. \eqref{Quasiperiodicity}, the antiperiodic case. 

The spinor vacuum free energy was previously reduced to evaluating the heat kernel function \eqref{Local representation} of the operator $\slashed{\nabla}^2(m)$, characterized by the eigenfunctions in {Eq.~\eqref{Complete solution}} and the eigenvalues in {Eq.~\eqref{Eigenvalues}}. Substituting these into {Eq.~\eqref{Local representation}} results in the diagonal matrix {\eqref{Matrix heat kernel}}, whose spinor components $K_s$ are defined in {Eq.~\eqref{Integral}} in terms of the integral $\mathcal{I}_s(r, q, \kappa)$. Since the order of the Bessel function depends on the continuous quantum number $\nu$ for $\kappa \neq 0$, the evaluation of $\mathcal{I}_s(r, q, \kappa)$ can become quite complex. To bypass an extensive and complex calculation that lies beyond our main objective, a more tractable expression for the spinor components is derived in the Appendix \ref{Heat kernel}. The final result is presented as follows
\begin{align}\label{Ks SD}
  K_{s} &= \frac{e^{-\tau m^2}}{(4\pi \tau)^{3/2}} \left[\sum_{\ell} (-1)^{\ell} e^{-\frac{ i\pi \ell s}{q}} e^{-\frac{\Delta \zeta_{\ell}^2}{4\tau}}    \right.\nonumber\\
  -&\left. \frac{q}{\pi^2} \sum_{n=-\infty}^{\infty} (-1)^n e^{\frac{ i\pi n s}{q}}  \int_{0}^{\infty} dy \,e^{-\frac{\Delta \zeta_{n,y}^2}{4\tau}} M_{n,q}(\Delta \phi, y)  \right] . 
\end{align}
 Here, we introduced the quantities $ \Delta \zeta_{\ell}^2$ and $ \Delta \zeta_{n,y}^2$, which are explicitly defined in the form
\begin{align}\label{Delta Zeta}
    &\Delta \zeta_{\ell}^2 = \rho^2 + \rho'^2 -2\rho\rho' \cos\left(\frac{2\pi \ell}{q} -\Delta\phi\right) + (\Delta Z - \bar{\kappa} \ell )^2,\nonumber\\
    &\Delta \zeta_{n,y}^2 = \rho^2 + \rho'^2 +2\rho\rho' \cosh(y) + (\Delta Z + \bar{\kappa} n)^2, 
\end{align}
with ${\bar{\kappa} = 2\pi \kappa /q}$. Moreover, the summation over $\ell$ is performed under the same condition as specified in {Eq.~\eqref{Summation over l}}. The explicit form of the function $M_{n,q}(\Delta \phi, y)$ for the case ${\Delta \phi \neq 0}$ can be found in Appendix \ref{Heat kernel}, although it is not required for the analysis presented here. In the coincidence limit, where ${\boldsymbol{r}' \rightarrow \boldsymbol{r}}$, it takes the form
\begin{align}
     M_{n,q}(0, y) = \frac{\frac{q}{2}+n}{\left(\frac{q}{2}+n\right)^2 +\left(\frac{q y}{2\pi}\right)^2}.
\end{align}
Furthermore, in this limit, the substitution of the components $K_s$ from \eqref{Ks SD} into {Eq.~\eqref{Matrix heat kernel}} reduces the spinor heat kernel $\boldsymbol{K}_{\slashed{\nabla}^{2}(m)}(\boldsymbol{r}, \boldsymbol{r}, \tau)$ to the same form as Eq. \eqref{Local HQ CS}, namely
\begin{align}\label{Local HQ SD}
    \boldsymbol{K}_{\slashed{\nabla}^{2}(m)}(\boldsymbol{r}, \boldsymbol{r}, \tau) &= \boldsymbol{K}^{\mathbb{E}}(\boldsymbol{r}, \boldsymbol{r}, \tau) \nonumber\\
     &+ \text{diag}(K^{\mathbb{T}}_{+}, K^{\mathbb{T}}_{-}, K^{\mathbb{T}}_{+}, K^{\mathbb{T}}_{+}).
 \end{align}
The first term corresponds to the standard Euclidean local heat kernel contribution, as given by {Eq.~\eqref{Euclidean C}}. The second term, represented by the components ${K^{\mathbb{T}}_{s} = K^{\mathbb{T}}_{s}(\boldsymbol{r}, \boldsymbol{r}, \tau)}$, arises due to the spacetime nontrivial topology. This contribution is expressed as follows
\begin{align}
   K^{\mathbb{T}}_{s} &= \frac{e^{-\tau m^2}}{(4\pi \tau)^{3/2}} \left[2\sum_{\ell=1}^{[q/2]}\,(-1)^{\ell }\cos\left(\ell \pi s/q\right) e^{-\frac{(\ell \bar{\kappa})^2 +(2\rho u_{\ell})^2}{4\tau}} \right.\nonumber\\ 
  &-\frac{q}{\pi^2}\sum_{n = - \infty}^{\infty} (-1)^n e^{\frac{ i\pi n s}{q}} \times \nonumber\\
 &\times \left. \int_{0}^{\infty}dy \,e^{-\frac{(n \bar{\kappa})^2 +(2\rho u_{y})^2}{4\tau}}\,M_{n,q}(0, y)\right] . 
\end{align}

Taking the trace over the spinor indices in {Eq.~\eqref{Local HQ SD}} and integrating over the whole space yields the global spinor heat kernel. This procedure directly leads to the asymptotic behavior of the heat kernel for small $\tau$ in the case ${\kappa = 0}$, as described in \eqref{Asymptotic expansion}.
Here, for comparison purposes, it is helpful to preserve the same structure, which is as follows
%Now, by taking the trace over the spinor indices in {Eq.~\eqref{Local HQ SD}}, and integrating in the whole space, we arrive at the global spinor heat kernel:
%\begin{align}
 %   &  K_{\slashed{\nabla}^2(m)}(\tau) = \frac{e^{-\tau m^2}}{(4\pi \tau)^{3/2}} \times \nonumber\\
  %  &\times \left[4V_3 + \frac{2\pi V_1}{q}\left(2\sum_{\ell=1}^{[q/2]}\,(-1)^{\ell }\cos\left(\ell \pi/q\right) \frac{e^{-\frac{(\ell \bar{\kappa})^2}{4\tau}}}{u_\ell^2} \right. \right.\nonumber\\ 
 %&-\left.\left.\frac{q}{\pi^2}\sum_{n = - \infty}^{\infty} \int_{0}^{\infty}dy \,\frac{e^{-\frac{(n \bar{\kappa})^2}{4\tau}}}{u_y^2}\,\mathcal{M}_{n,q}(0, y)\right) \tau\right] ,
%\end{align}
\begin{align}\label{c0 C1}
 K_{\slashed{\nabla}^2(m)}(\tau)  =   \frac{e^{-\tau m^2}}{\left( 4\pi \tau \right)^{3/2}} [c_0 + c_{1}(\tau)\tau]  .
\end{align}
While $c_0$ is a heat kernel coefficient that retains the same form as in {Eq.~\eqref{c_0}}, the function $c_1(\tau)$ does not strictly qualify as one, despite the above structure suggesting a small-$\tau$ asymptotic expansion.  In addition to the parameters related to the topology of the cosmic dispiration spacetime (CD), it explicitly depends on the parameter $\tau$, given by the following expression
\begin{align}\label{c1(tau)}
    c_1(\tau) &= R_{\text{CD}}(q, \kappa) V_1\nonumber\\
    &=\frac{8\pi}{q}\left[\sum_{\ell=1}^{[q/2]}\,(-1)^{\ell }\cos\left(\ell \pi/q\right) \frac{e^{-\frac{(\ell \bar{\kappa})^2}{4\tau}}}{u_\ell^2} \right.\nonumber\\ 
  &-\left.\frac{q}{2\pi^2}\int_{0}^{\infty}\frac{dy}{u_y^2} \sum_{n = - \infty}^{\infty} e^{-\frac{(n \bar{\kappa})^2}{4\tau}}\,\mathcal{M}_{n,q}(0, y)\right]  V_1 ,
\end{align}
with
\begin{align}\label{Function mathcal M}
    \mathcal{M}_{n,q}(0, y) = (-1)^n \cos(n \pi/q)\,M_{n,q}(0, y).
\end{align}
For small $\tau$, all terms within the brackets are exponentially suppressed, except for the first term and the contribution associated with ${n=0}$ in the summation over $n$. Therefore, we have
%Although the structure of the expression \eqref{c0 C1} suggests an asymptotic behavior of the heat kernel at small $\tau$, $c_1(\tau)$ does not strictly qualify as a heat kernel coefficient due to its dependence on the parameter $\tau$. 
%However, when evaluated at small $\tau$, all terms within the brackets are exponentially suppressed, except for the contribution associated with ${n=0}$ in the summation over $n$.
%The heat kernel coefficient $c_1$ corresponds to the leading contribution of $c_1(\tau)$ when evaluated for small $\tau$. In fact, in this regime, all terms within brackets are exponentially suppressed, except for the contribution associated with ${n=0}$ in the summation over $n$. Therefore, we have
\begin{align}
    c_1(\tau \rightarrow 0) = -\frac{4V_1}{\pi} \int_{0}^{\infty} dy\, \frac{ \mathcal{M}_{0,q}(0, y)}{\cosh^{2}(y/2)}  + \mathcal{O}(e^{-1/\tau}) ,
\end{align}
where $\mathcal{O}(e^{-1/\tau})$ stands for those terms going to zero faster than any positive power of $\tau$ and, therefore, can be neglected. 
The heat kernel coefficient $c_1$ thus corresponds to the leading contribution of $c_1(\tau)$ in the small-$\tau$ regime, and is given by
\begin{align}\label{c1(q) CD}
    c_1 = -\frac{4V_1}{\pi} \int_{0}^{\infty} dy\, \frac{ \mathcal{M}_{0,q}(0, y)}{\cosh^{2}(y/2)} \simeq 4\left(- V_1\frac{\pi}{3q} \right) .
\end{align}
Consequently, the spinor heat kernel $K_{\slashed{\nabla}^2(m)}(\tau)$ in cosmic dispiration spacetime admits an expansion in powers of $\tau$, analogous to that presented in Eq.~\eqref{heat equation expansion}, featuring only two nonvanishing heat kernel coefficients
\begin{align}\label{Local heat kernel expansion}
 K_{\slashed{\nabla}^2(m)}(\tau)  \sim   \frac{e^{-\tau m^2}}{\left( 4\pi \tau \right)^{3/2}} (c_0 + c_1 \tau) + \mathcal{O}(e^{-1/\tau}) . 
\end{align}

\begin{figure}
    \centering
    \includegraphics[width=0.47\textwidth]{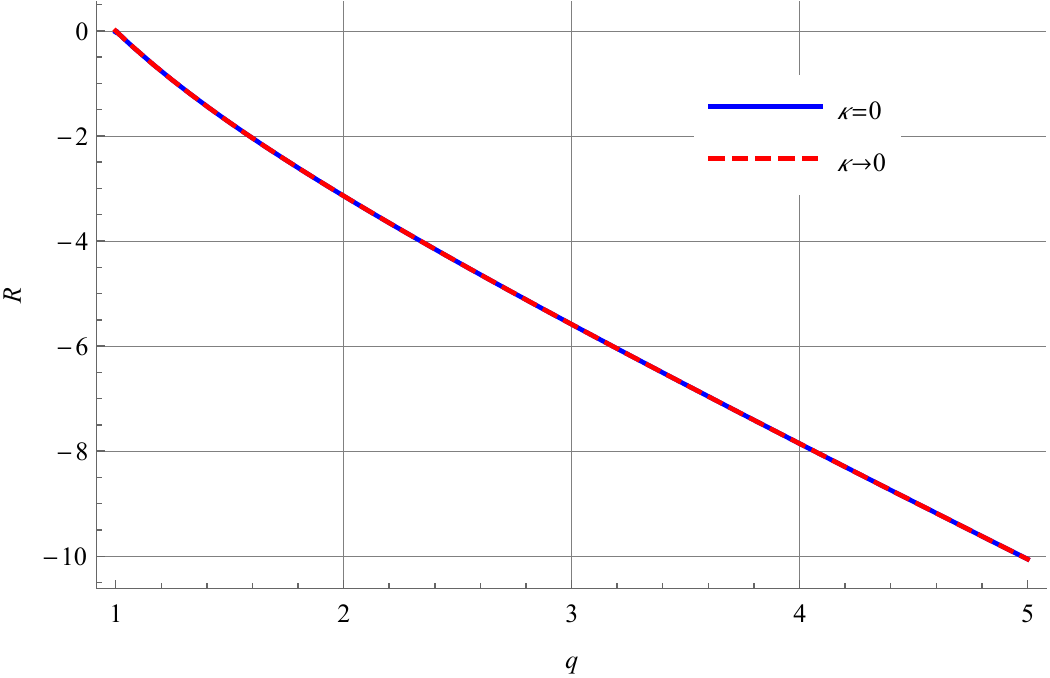}\\
    \caption{The plot compares the ratio ${R=c_1/V_1}$ as a function of $q$ in cosmic dispiration spacetime in limiting case ${\kappa \rightarrow 0}$ with its equivalent in cosmic string spacetime (${\kappa = 0}$).
    In the case of the continuous curve, $R$ corresponds to Eq. \eqref{c1(tau)} in the limit ${\kappa \rightarrow 0}$. Conversely, for the dashed curve, it represents Eq. \eqref{c1(q)}. Notably, the two curves align perfectly, demonstrating complete overlap.}
    \label{Plot c1 2}
\end{figure}

In the case ${\kappa = 0}$ discussed in the previous section, corresponding to the pure cosmic string, the global factor that relates the spinor coefficient $c_1(q,0)$ in {Eq.~\eqref{c1(q)}} with the corresponding scalar coefficient is ``${-2}$" in four dimensions. When ${\kappa \neq 0}$, this factor becomes $4$, as evident in {Eq.~\eqref{c1(q) CD}}, coinciding with the value of $c_0$. This modification in the structure of $c_1$ in \eqref{c1(q) CD} results from the inclusion of the parameter $\kappa$, associated with screw dislocation, which endows the spacetime with a different topology. Interestingly, it remains independent of $\kappa$, although its form arises due to the condition ${\kappa \neq 0}$.
The heat kernel function is expressed in the form of {Eq.~\eqref{c0 C1}} to allow a direct comparison with the corresponding function obtained for the case ${\kappa = 0}$, which represents the pure cosmic string.
In the limiting case ${\kappa = 0}$, Eq. \eqref{c0 C1} exactly reproduces the asymptotic behavior of the heat kernel at small $\tau$. This establishes $c_1(\tau)$, now independent of $\tau$, as a heat kernel coefficient. Furthermore, its structure closely resembles the heat kernel coefficient presented in {Eq.~\eqref{c_1}} for ${\chi = 0}$, except for the summation over $n$ of the function $\mathcal{M}_{n,q}(0, y)$. 
This series is found to converge to a complex expression involving hypergeometric functions, making direct analytical comparisons with previous results quite challenging. Instead, we numerically analyze the behavior of $H_{\text{CD}}(q,\kappa)$ in ${\kappa = 0}$ and compare it with the analytical expression for $H_{\text{CS}}(q, \chi)$ from {Eq.~\eqref{c_1}} for ${\chi = 0}$.
The comparison is depicted in {Fig.~\ref{Plot c1 2}}, where the two curves exhibit a notable alignment, completely overlapping. Consequently, for ${\kappa = 0}$, the heat kernel coefficient of {Eq.~\eqref{c1(q) CD}} faithfully recovers the structure described in {Eq.~\eqref{c1(q)}}.
With this established correspondence between the heat kernel functions and their associated heat kernel coefficients, all results derived for the cosmic dispiration spacetime can be reduced to those of the cosmic string spacetime as a special case.

As seen in the cosmic string case, nonvanishing heat kernel coefficients contribute to terms in the zeta function, and subsequently in the spinor vacuum energy, which increase with nonnegative powers of the mass. To analyze such terms, it is convenient to express the heat kernel function in the following form
\begin{align}\label{Heat kernel CD}
    &  K_{\slashed{\nabla}^2(m)}(\tau) =
     \frac{e^{-\tau m^2}}{\left( 4\pi \tau \right)^{3/2}} (c_0 + c_1 \tau)\nonumber\\
   & +\frac{8\pi V_1}{q} \frac{e^{-\tau m^2}}{(4\pi \tau)^{3/2}}\left(\sum_{\ell=1}^{[q/2]}\,(-1)^{\ell }\cos\left(\ell \pi/q\right) \frac{e^{-\frac{(\ell \bar{\kappa})^2}{4\tau}}}{u_\ell^2} \right. \nonumber\\ 
  &-\left.\frac{q}{2\pi^2} \int_{0}^{\infty}\frac{dy}{u_y^2}  \,\mathop{{\sum}'}\limits_{n=-\infty}^{\infty}e^{-\frac{(n \bar{\kappa})^2}{4\tau}}\,{\mathcal{M}}_{n,q}(0, y)\right) \tau .
\end{align}
where the prime in the summation over $n$ denotes that the term $n= 0$ should be withdrawn from the series.
The corresponding zeta function is found to be
\begin{align}
  &\,\Gamma\left(w- 1/2\right) \zeta_{\slashed{\nabla}^2(m)}\left(w- 1/2\right)\nonumber\\
      & =  \frac{1}{(4\pi)^{3/2}}\sum_{p =0}^{1} \frac{c_p\,\Gamma\left(w + p - 2 \right)}{ \left(m^2\right)^{w + p - 2}} \nonumber\\
      &+ \frac{2^{2-w}V_1}{\sqrt{\pi} q} m^{1-w}\left[  \sum_{\ell=1}^{[q/2]}\,(-1)^{\ell } \frac{\cos\left(\ell \pi/q\right)}{u_\ell^2} \frac{K_{1-w}(|\ell m \bar{\kappa}|)}{(\ell \bar{\kappa})^{1-w}} \right. \nonumber\\
     &-\left.\frac{q}{2\pi^2}\int_{0}^{\infty}\frac{dy}{u_y^2} \, \mathop{{\sum}'}\limits_{n=-\infty}^{\infty}\mathcal{M}_{n,q}(0, y) \frac{K_{1-w}(|n m \bar{\kappa}|)}{(n \bar{\kappa})^{1-w}} \right] , 
\end{align}
which contains simple poles at ${w = 0}$, with residue involving coefficients $c_0$ and $c_1$ that increase with mass $m$ without bounds, {Eq.~\eqref{Residue}}. 
Since these coefficients are nonvanishing, the generalized zeta function is not finite. Therefore, as previously argued, all contributions from $c_0$ and $c_1$ must be removed to ensure that the spinor vacuum energy satisfies the renormalization condition \eqref{renormalization condition}. By doing so, the Casimir energy as given in {Eq.~\eqref{Renormalized Casimir energy}} per unit length $V_1$, associated with a massive spinor field in a cosmic dispiration spacetime, is expressed as follows
\begin{align}\label{Massive CE}
  \mathcal{E}_{\text{C}}^{\text{ren}}(m) &= \frac{1}{\sqrt{\pi} q} \left[  \sum_{\ell=1}^{[q/2]}\,(-1)^{\ell }\,\frac{\cos\left(\ell \pi/q\right)}{u_\ell^2} \frac{m\,K_{1}(m \ell\bar{\kappa})}{\ell \bar{\kappa}} \right. \nonumber\\
     &-\left.\frac{q}{2\pi^2}\int_{0}^{\infty}\frac{dy}{u_y^2}  \mathop{{\sum}'}\limits_{n=-\infty}^{\infty}\mathcal{M}_{n,q}(0, y)\frac{m\,K_{1}(m n \bar{\kappa})}{n \bar{\kappa}} \right] .
\end{align}
When ${m \bar{\kappa} \gg 1}$, the Macdonald function exhibits the asymptotic behavior ${K_\mu(x) \simeq \sqrt{\frac{\pi}{2x}}\,e^{-x}}$. In this limiting case, the Casimir energy density decreases exponentially as the field's mass increases, thus satisfying the renormalization condition \eqref{renormalization condition}. This result originates from the fact that quantum fluctuations and, hence, Casimir energy are absent for an infinitely massive field \cite{Bordag2000}. 

\begin{figure} 
    \centering
    \hspace{-7.2cm}$(a)$\\
    \includegraphics[width=0.47\textwidth]{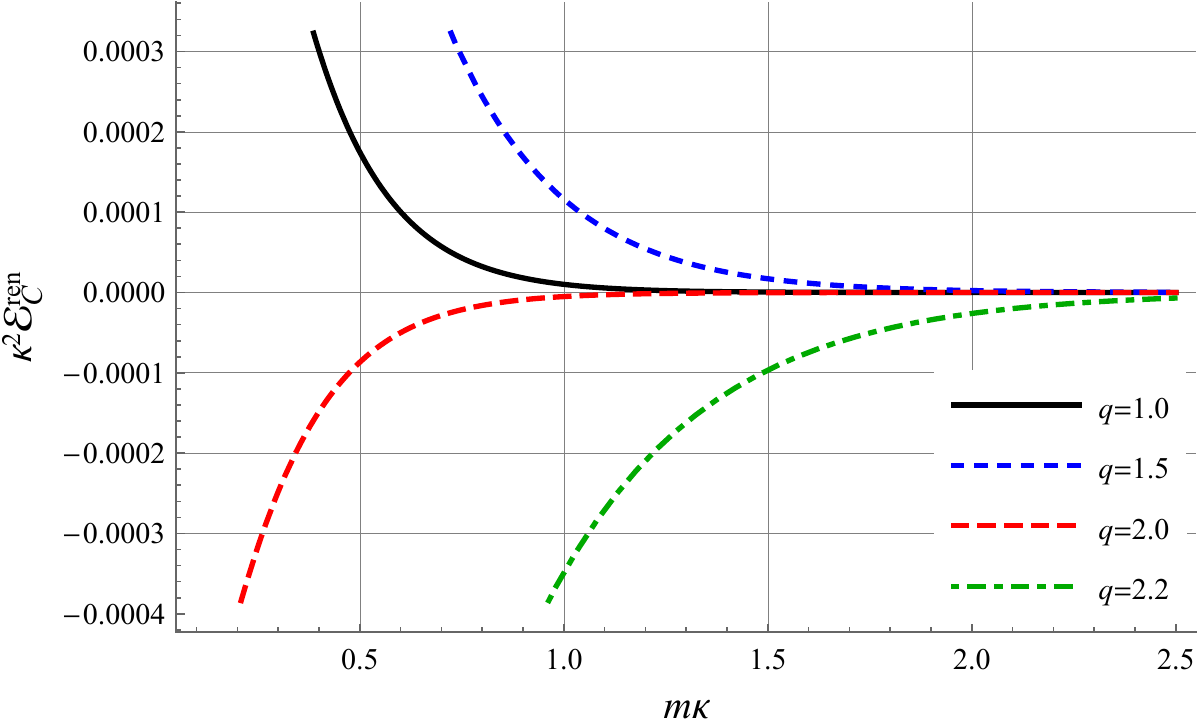}\\
     \hspace{-7.2cm}$(b)$\\
    \includegraphics[width=0.47\textwidth]{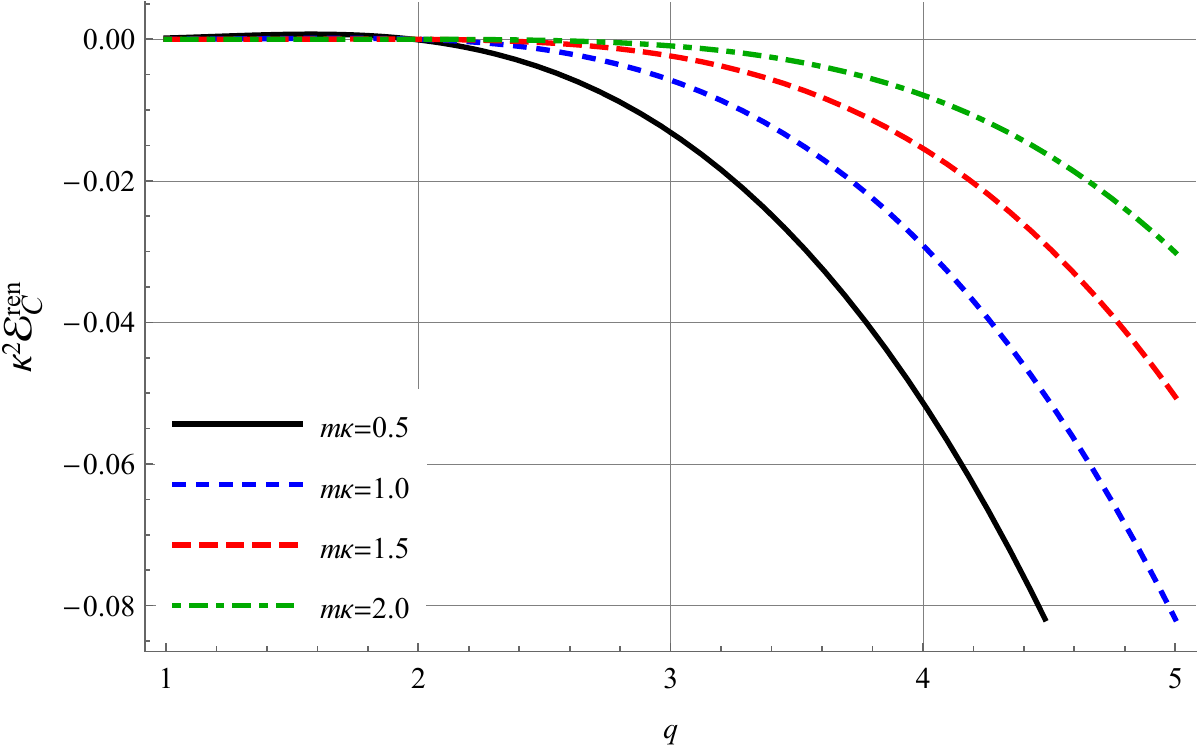}
   \caption{$(a)$ displays the vacuum energy density as a function of $m \kappa$, and $(b)$ presents it in terms of $q$. Both plots illustrate variations corresponding to different parameter values.}
    \label{Plot Energy}
\end{figure}

%At this stage, it is important to emphasize that adopting the renormalization scheme by subtraction can lead to a negative vacuum energy density. 
 {Figure~\ref{Plot Energy}}$(a)$ displays the above vacuum energy density as a function of $m \kappa$, while {Fig.~\ref{Plot Energy}}$(b)$ presents it in terms of $q$. Both plots illustrate variations corresponding to different parameter values. Notably, as $m\kappa$ increases, the vacuum energy density decays exponentially to zero, as expected.
Observe that the vacuum energy density is positive for certain values of $q$; see, for instance, the curve for ${q = 1}$ in {Fig.~\ref{Plot Energy}}$(a)$, which only accounts for the contribution of screw dislocation. In contrast, for other values of $q$, the vacuum energy density is negative, as depicted by the curves corresponding to ${q = 2}$ and ${q = 2.2}$ in {Fig.~\ref{Plot Energy}}$(a)$, as well as the curves shown in {Fig.~\ref{Plot Energy}}$(b)$. In \cite{Ford2010}, these negative values are referred to as subvacuum phenomena, arising directly from the renormalization scheme by subtraction. Furthermore, some proposals are discussed to measure these effects.

%At this stage, it is important to emphasize that adopting the renormalization scheme by subtraction can lead to a negative vacuum energy density. Indeed, for certain values of $q$, the vacuum energy density becomes negative, as depicted by the curve corresponding to ${q = 2}$ in part $(a)$ of {Fig.~\ref{Plot Energy}}. These negative values are examples of sub-vacuum phenomena, as discussed in {Ref.~\cite{Ford2010}}.

The vacuum energy density in the massless case is derived from the asymptotic behavior of the Macdonald function with small arguments, as defined in {Eq.~\eqref{Small arguments limit}}, following an approach analogous to that in the previous section. By applying the same steps, one can establish that the following massless limit holds
\begin{align}\label{Massless CE}
  \mathcal{E}_{\text{C}}^{\text{ren}}(m) &= \frac{q}{4 \pi^3 \kappa^2} \left[  \sum_{\ell=1}^{[q/2]}\,(-1)^{\ell }\,\frac{\cos\left(\ell \pi/q\right)}{\ell^2 u_\ell^2}  \right. \nonumber\\
     &-\left.\frac{q}{2\pi^2}\int_{0}^{\infty}\frac{dy}{u_y^2} \mathcal{G}_{2}(q, y) \right] ,
\end{align}
where the function $ \mathcal{G}_{\nu}(q, y)$ is expressed as follows
\begin{align}
    \mathcal{G}_{\nu}(q, y) = \mathop{{\sum}'}\limits_{n=-\infty}^{\infty} \frac{\mathcal{M}_{n,q}(0, y)}{n^\nu},
\end{align}
with $\mathcal{M}_{n,q}(0, y)$ defined in {Eq.~\eqref{Function mathcal M}}. 

As discussed previously, since ${c_2(m)}$ is identically null in the massless case, $\zeta_{\slashed{\nabla}^2(m)}\left(-1/2\right)$ remains finite, rendering the FP prescription redundant in the Casimir energy. This makes the renormalization procedure unnecessary, as all ambiguities are naturally eliminated and the renormalization condition \eqref{renormalization condition} is trivially satisfied.
It is also important to emphasize that the Casimir energy densities obtained in this section are strictly valid for ${\kappa \neq 0}$. 
In particular, the contribution purely associated with the screw dislocation parameter $\kappa$ is achieved by setting ${q = 1}$, in which case the summation over $\ell$ vanishes. In the limiting case ${\kappa \rightarrow \infty}$, this contribution vanishes in the massless regime.

\vspace{.5cm}
\section{Finite-temperature corrections} \label{S6}

Now, we have the condition to investigate the temperature corrections to the Casimir energy density. 
To determine the structure of thermal corrections for the Casimir energy, as defined in {Eq.~\eqref{Generic temperature correction}}, one must evaluate an integral over $\tau$ involving the global heat kernel \eqref{Heat kernel CD}. In doing so, the first two terms, associated with the coefficients $c_0$ and $c_1$, reproduce the expressions given in Eqs. \eqref{Euclidean TC} and \eqref{Topological TC} precisely. In particular, in the massless limit, these contributions follow a power-law behavior in $T$, {Eq.~\eqref{Massless TC}}, where the term containing $c_0$ represents the spinor black-body radiation, as given in {Eq.~\eqref{Alpha coefficients}}.

As we have seen, nonzero heat kernel coefficients signal the presence of divergences, and terms that associate them must be subtracted. The resulting renormalized temperature correction term associated with a massive spinor field is expressed in terms of a double sum as follows
\begin{align}\label{CD TC 2}
  &\Delta F^{\text{ren}}(\beta, m) = \frac{2}{q\pi}V_1 \sum_{n=1}^{\infty} \cos(\pi n)\times \nonumber\\
  &\times \left[  \sum_{\ell=1}^{[q/2]}\,(-1)^{\ell }\,\frac{\cos\left(\ell \pi/q\right)}{u_\ell^2} \frac{m\,K_{1}(m R_{n,\ell })}{  R_{n,\ell }} \right. \nonumber\\
     &-\left.\frac{q}{2\pi^2}\int_{0}^{\infty}\frac{dy}{u_y^2} \mathop{{\sum}'}\limits_{k=-\infty}^{\infty}\mathcal{M}_{k,q}(0, y) \frac{m\,K_{1}(m R_{n, k})}{ R_{n, k}} \right] ,
\end{align}
where $R_{n,b}$ is given by 
\begin{align}\label{RNB}
    R_{n,b} = \sqrt{(n\beta)^2 + (b\bar{\kappa})^2} , \quad b = \ell, k. 
\end{align}
\begin{figure}
    \centering
    \hspace{-2.8cm}$(a)$ \qquad \qquad  \quad \qquad \qquad \qquad {\scriptsize${q=2.5}$}\\
    \includegraphics[width=0.46\textwidth]{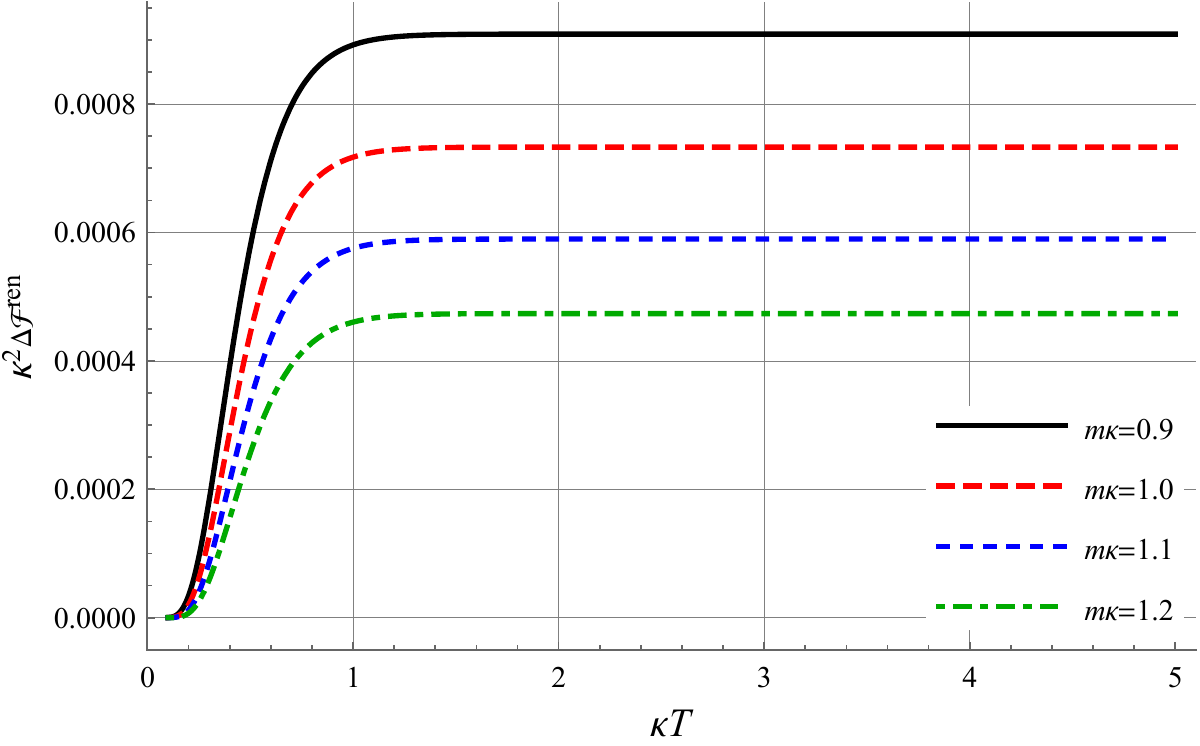}\\
     \centering
   \hspace{-2.8cm}$(b)$ \qquad \qquad  \quad \qquad \qquad \qquad {\scriptsize${q=1.0}$}\\
    \includegraphics[width=0.48\textwidth]{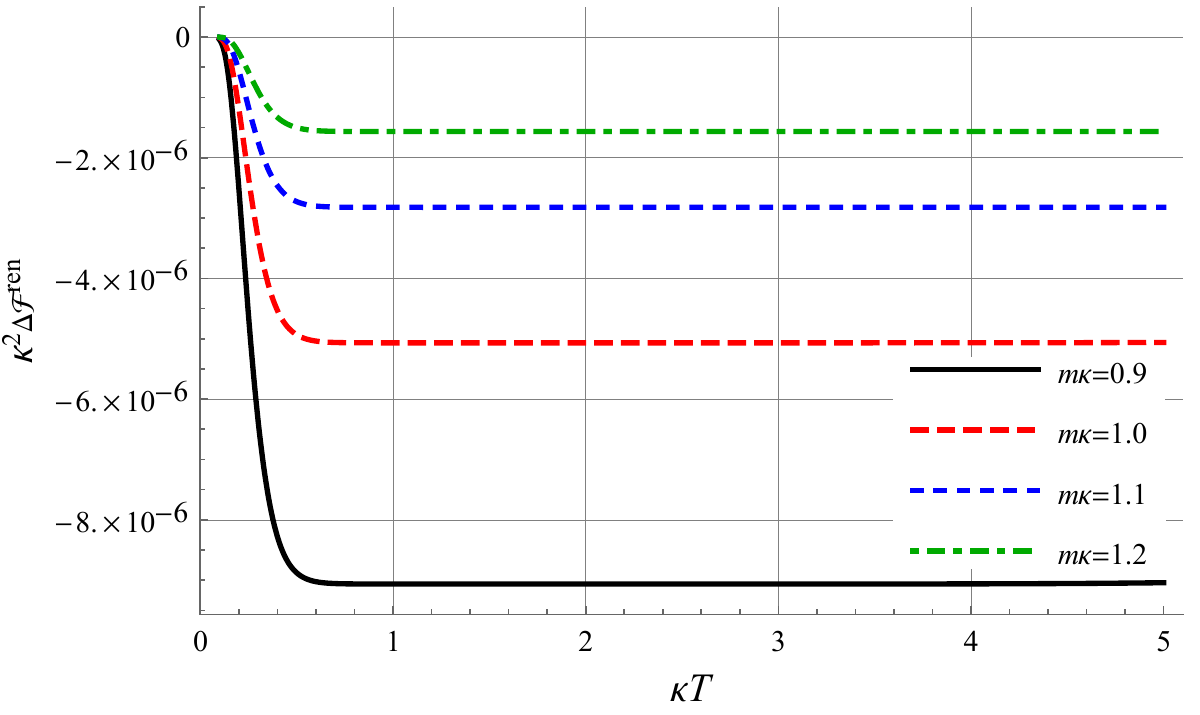}
   \caption{The renormalized temperature correction term per unit of $V_1$ is presented as a function of $\kappa T$ for different values of $m\kappa$, while maintaining $q$ fixed at two distinct values. $(a)$ considers $q=2.5$, while $(b)$ examines $q=1$, representing the contribution exclusively associated with the screw dislocation.}
    \label{Plot Correction}
\end{figure}

In particular, applying the limit \eqref{Small arguments limit} to the above temperature correction expression yields the massless contribution
\begin{align}\label{Massless TC2}
  \Delta F^{\text{ren}}(\beta) &= \frac{2}{q\pi}V_1 \sum_{n=1}^{\infty} \cos(\pi n)\times \nonumber\\
  &\times\frac{2}{q\pi } \left[  \sum_{\ell=1}^{[q/2]}\,(-1)^{\ell }\,\frac{\cos\left(\ell \pi/q\right)}{u_\ell^2 R_{n,\ell}^2}  \right. \nonumber\\
     &-\left.\frac{q}{2\pi^2}\int_{0}^{\infty}\frac{dy}{u_y^2} \mathop{{\sum}'}\limits_{k=-\infty}^{\infty}\frac{\mathcal{M}_{k,q}(0,y)}{R_{n,k}^2} \right] .
\end{align}

In Fig. \ref{Plot Correction}, we present the renormalized temperature correction term per unit of $V_1$ as a function of $\kappa T$ for various values of $m\kappa$ while keeping $q$ fixed at two distinct values. 
In each case, the plot shows that $\Delta F^{\text{ren}}$ approaches zero as $T$ tends to zero, as expected, while remaining independent of $T$ in the high-temperature limit.

\subsection{High-temperature limit}

Let us analyze the high-temperature limit, ${\beta \ll \bar{\kappa}}$ (or equivalently ${\bar{\kappa}T \gg 1}$), of the
final expression \eqref{Massless TC2}. This can be achieved by performing the summation over $n$ first, which yields 
\begin{align}
   \frac{1}{\beta^2} \sum _{n=1}^{\infty } \frac{\cos (\pi  n)}{\left[ n^2+\left(\frac{b \bar{\kappa }}{\beta}\right)^2\right]}= -\frac{1}{2 b^2 \bar{\kappa }^2}+\frac{\pi  T}{2 b \bar{\kappa }}\,\text{csch}\left(b \pi \bar{\kappa } T\right) .
\end{align}
Here, $b$ is defined as in \eqref{RNB}, representing the index of the remaining summation in \eqref{Massless TC2}, which can be either $\ell$ or $k$. By evaluating the above series in the limit ${\bar{\kappa} T \gg 1}$ and substituting it into Eq. \eqref{Massless TC2}, we are left with the following expression
\begin{align}\label{MRTC H}
    \Delta F^{\text{ren}}(\beta \ll \bar{\kappa}) & \simeq  -\frac{q V_1}{4 \pi ^3 \kappa ^2} \left[\sum _{\ell =1}^{[q/2]} \frac{(-1)^{\ell} \cos \left(\frac{\pi \ell}{q}\right)}{\ell^2 u_\ell^2} \right.\nonumber\\
   &-\left.\frac{q}{2 \pi ^2}\int _0^{\infty }\frac{dy}{u_y^2}\,\mathcal{G}_2(y,q)\right] +\mathcal{O}(e^{-\pi \bar{\kappa} T}) ,
\end{align}
where $\mathcal{O}(e^{-\pi \bar{\kappa} T})$ represents exponentially suppressed terms at high $T$. As these terms vanish in the limit $T \to \infty$, 
the above expression does not depend on the temperature, consistent with the behavior shown in Fig. \ref{Plot Correction} for ${\kappa T\gg 1}$.
Moreover, it corresponds precisely to the Casimir energy for the massless spinor field, {Eq.~\eqref{Massless CE}}, but with the opposite sign. 
Therefore, the effect of this temperature correction term entirely compensates the spinor Casimir energy within the free energy \eqref{Reg free energy}. Such cancellations constitute an intrinsic characteristic of temperature corrections associated with spinor fields at the high-temperature limit \cite{Plunien1986, Joas2024}.

Note the physical significance of subtracting the contributions associated with the coefficients $c_0$, which lead to the Stefan-Boltzmann law, proportional to $T^4$, and $c_1$, proportional to $T^2$, in achieving the correct classical limit. Unlike scalar fields, spinor fields lack a temperature correction term that varies linearly with $T$, resulting in a spinor free energy that is renormalized to zero at very high temperatures.

\subsection{Low-temperature limit}

We now examine the asymptotic expansion of Eq.~\eqref{Massless TC2} in the low-temperature regime, characterized by ${\beta \gg \bar{\kappa}}$, or equivalently ${\bar{\kappa} T\ll 1}$. For the first term inside the brackets, we begin by evaluating the summation over $n$, yielding
 \begin{align}\label{Sum in n}
   &\frac{1}{\beta^2} \sum _{n=1}^{\infty } \frac{1}{n^2}\frac{\cos (\pi  n)}{\left[ 1+\left(\frac{\ell \bar{\kappa }}{n\beta}\right)^2\right]} \nonumber\\
   &=  \sum_{p=0}^{\infty}(-1)^p \left[2^{-(1+2p)} -1\right] \zeta(2p+2)\,\left( \ell \bar{\kappa} T\right)^{2p}. 
\end{align}
We first apply a binomial expansion for ${\ell \bar{\kappa} T\ll 1}$. In the resulting expression, we perform the summation over $n$, separating it into contributions from even and odd terms. As discussed in subsection \ref{kappa zero}, this process leads to the appearance of the Riemann and Hurwitz zeta functions, which are related through {Eq.~\eqref{HZF relation}} and ultimately produce the above result.

For the second term inside the brackets, it is more appropriate to first perform the summation over $k$, which converges to an extensive and complex expression involving hypergeometric functions. Expanding the result for ${\frac{\bar{\kappa}}{n\beta} \ll 1}$ and then performing the sum in $n$ for the leading term of the expansion yields the following
 \begin{align}\label{Sum in k}
   &\frac{1}{\bar{\kappa}^2}\sum_{n=1}^{\infty} \cos(\pi n) \mathop{{\sum}'}\limits_{k=-\infty}^{\infty}\frac{\mathcal{M}_{k,q} (0, y)}{\left[k^2+\left(\frac{n\beta}{ \bar{\kappa }}\right)^2\right]} \nonumber\\
   & =   \left[\frac{2\pi}{q} \frac{1}{\pi^2 +y^2}-\sum_{k=-\infty}^{\infty}\mathcal{M}_{k,q} (0, y)  \right] \frac{T^2}{2}\zeta(2) \nonumber\\
   &+\mathcal{O}(T^{2p}), \quad p \geq 2.  
\end{align}
By inserting Eqs. \eqref{Sum in n} and \eqref{Sum in k} into {Eq.~\eqref{Massless TC2}} and considering terms up to order $T^2$, we obtain the following asymptotic behavior for the low-temperature limit
\begin{align}\label{MRTC L}
  \Delta F^{\text{ren}}(\beta \gg \bar{\kappa}) &\simeq -\left[R_{\text{CD}}(q,0)+\frac{1}{3q}\right]\frac{V_1}{8 \pi ^2}T^2 \zeta (2) .
\end{align}
The function $R_{\text{CD}}(q,0)$ is defined in \eqref{c1(tau)} and satisfies ${R_{\text{CD}}(1,0)= 0}$. 
This implies that even when the deficit angle is absent, ${q=1}$, there remain contributions originating from the screw dislocation. Moreover, the above expression approaches zero as the temperature nears zero, in accordance with the behavior depicted in {Fig.~\ref{Plot Correction}} for ${\kappa T \ll 1}$. Hence, in the very low-temperature limit, the spinor free energy is dominated by the Casimir energy density at zero temperature, as expected  \cite{Basil1978}.

\section{Conclusion}\label{Conclusion}

%In this work, we have investigated the finite temperature Casimir effect using the generalized zeta function method for a massive spinor field defined on topologically nontrivial spacetime around a chiral cosmic string, the cosmic dispiration spacetime.

In this study, we consider a massive spinor field in spacetime resulting from the combination of a cosmic string and a screw dislocation, referred to as cosmic dispiration. The nontrivial topology of this spacetime modifies the quantum modes of the spinor field in the vacuum state, giving rise to the Casimir effect. We analyze this phenomenon at both zero and finite temperatures using the generalized zeta function method, which offers a robust framework for investigating divergences associated with nonzero heat kernel coefficients and removing them in the renormalization scheme by subtraction.
%Consequently, their renormalized expressions, {Eq.~\eqref{Massive CE}} and {Eq.~\eqref{CD TC 2}}, are both nonvanishing.  The zero-temperature Casimir energy density, as depicted in {Fig.~\ref{Plot Energy}}, exhibits either positive or negative values depending on the choice of $q$ and decreases exponentially as the field’s mass increases.

Before examining cosmic dispiration spacetime, we first analyze the limiting case of the cosmic string spacetime endowed with a quasi-antiperiodic boundary condition. In this case, both the Casimir energy and its temperature corrections depend on the two nonzero heat kernel coefficients, $c_0$ and $c_1(q, \chi)$, vanishing when renormalized as shown in Eqs. \eqref{RCE C1} and \eqref{RTE C1}. 
The coefficient $c_0$ corresponds to the Euclidean divergence, {Eq.~\eqref{c_0}}, while $c_1(q, \chi)$ represents the topological contribution combined with the quasi-antiperiodic condition, {Eq.~\eqref{c_1}}. Our results generalize those presented in {Ref.~\cite{Fursaev1997}} for spin-$1/2$ fields, both matching when ${\chi = 0}$. In particular, $c_1(q, 0)$ in {Eq.~\eqref{c1(q)}} is equal to minus twice the same coefficient obtained for the scalar field in {Ref.~\cite{Herondy2024}}, agreeing with the expected value in four dimensions \cite{Fursaev1997}. Furthermore, an intriguing aspect of the coefficient $c_1(q, \chi)$ is illustrated in {Fig.~\ref{Plot c1 1}}. By including quasi-antiperiodicity effects (${\chi \neq 0}$), it is possible to completely cancel out the topological contribution (${q>1}$), resulting in ${c_1(q, \chi) = 0}$ and leaving only the coefficient $c_0$ associated with the Euclidean divergence.

We then extend this analysis to the cosmic dispiration spacetime. Again, the only nonzero heat kernel coefficients are $c_0$, related to the Euclidean divergence, and $c_1(q)$, associated with the nontrivial topology divergence. Although the presence of screw dislocation endows the spacetime with different topology, the structure of $c_1(q)$ is modified so that it does not depend on the screw dislocation parameter $\kappa$, {Eq.~\eqref{c1(q) CD}}. Moreover, its value is four times greater than the corresponding scalar field coefficient presented in {Ref. \cite{Herondy2024}}.
These divergences are removed from the Casimir energy density and the corresponding temperature corrections in the renormalization by subtraction scheme. Consequently, their renormalized expressions, Eqs.~\eqref{Massive CE} and \eqref{Massless TC2}, are both nonvanishing. 
The renormalized Casimir energy density, as depicted in {Fig.~\ref{Plot Energy}}, exhibits either positive or negative values depending on the choice of the parameter $q$ encoding the deficit angle. Moreover, it decreases exponentially as the field’s mass increases.
We also analyzed the high- and low-temperature asymptotic limits of the renormalized temperature correction term in the massless regime, as expressed in Eqs. \eqref{MRTC H} and \eqref{MRTC L}, 
which are consistent with the curves illustrated in {Fig.~\ref{Plot Correction}}. The spinor vacuum free energy is renormalized to zero at very high temperatures and is dominated by the zero-temperature Casimir energy density at very low temperatures.

\section{Acknowledgments}

The authors are grateful to D. V. Vassilevich for the enlightening discussions. 
J. V. would like to thank Funda\c{c}\~{a}o de Amparo a Ci\^{e}ncia e Tecnologia do Estado de Pernambuco (FACEPE) for their partial financial support. A. M. acknowledges financial support from the Brazilian agencies Conselho Nacional de Desenvolvimento Científico e Tecnológico (CNPq), grant no. 306295/2023-7, and Coordenação de Aperfeiçoamento de Pessoal de Nível Superior (CAPES). H. M. is partially supported by CNPq under grant no. 308049/2023-3.

\begin{appendices}\label{Appendix}
\vspace{1cm}
\section{Evaluation of the integral $\mathcal{I}_s(r, q, \kappa)$} \label{Heat kernel}

In this section, we derive a more tractable expression for the components of the spinor heat kernel, as outlined in {Eq.~\eqref{Appendix 1}}. This requires evaluating the integral over the continuous quantum number $\nu$ presented in {Eq.~\eqref{Integral}}, namely
\begin{align}\label{Integral A}
    \mathcal{I}_s(r, q, \kappa) = \int_{-\infty}^{\infty}d\nu\,e^{-\tau \nu^2 + i\nu \Delta Z} \mathcal{S}_s(r, q, \kappa) , 
\end{align}
where $\mathcal{S}_s(r, q, \kappa)$ is a summation over the discrete quantum number $\ell$ given by
\begin{align}\label{Sum A}
    \mathcal{S}_s(r, q, \kappa) =  e^{\frac{is \Delta \phi}{2}} \sum_{\ell=-\infty}^{\infty} e^{i\mu_{s}(\ell, \nu)\Delta \phi}I_{|\mu_{s}(\ell, \nu)|}(r) .
\end{align}
The order of the Bessel function $\mu_{s}(\ell, \nu)$, previously defined in \eqref{Bessel order}, can be expressed in a more convenient form as follows
\begin{align}
    \mu_{s}(\ell, \nu) = q \left(\ell + \frac{1}{2} - h_s\right), \quad h_s = \frac{1}{q}\left(\kappa \nu + \frac{s}{2}\right). 
\end{align}

To evaluate the summation over $\ell$ in \eqref{Sum A}, we use the following integral representation of the modified Bessel function \cite{Abramowitz1972}
\begin{align}\label{Bessel Property}
    & I_{|\mu_{s}(\ell, \nu)|}(r)=\frac{1}{\pi }\int_0^{\pi }dy\, e^{r  \cos (y)} \cos \left(y |\mu_{s}(\ell, \nu)|\right) \nonumber\\
    &-\frac{ \sin \left(\pi  |\mu_{s}(\ell, \nu)|\right) }{\pi}\int_0^{\infty }dy\,e^{-|\mu_{s}(\ell, \nu)|y - r\cosh (y)} .
\end{align}
By substituting this into {Eq.~\eqref{Sum A}} and using the identity
\begin{align}\label{Delta 1}
\sum_{\ell= -\infty}^{\infty} e^{ibn} =  2\pi\sum_{\ell= -\infty}^{\infty} \delta(b-2\pi n) ,
\end{align}
particularly to the term involving the first integral in {Eq.~\eqref{Bessel Property}}, we arrive at the following result
\begin{align}
   \frac{1}{\pi } \sum_{\ell=-\infty}^{\infty} e^{i\mu_{s}(\ell, \nu)\Delta \phi}\int_0^{\pi }dy\, e^{r  \cos (y)} \cos \left(y |\mu_{s}(\ell, \nu)|\right) \nonumber\\
   = \frac{1}{q}\sum_{\ell}(-1)^\ell  e^{-2 \pi  i \ell h_s} e^{r \cos\left(\frac{2 \pi  \ell}{q}-\Delta \phi \right)}. 
\end{align}
The summation over $\ell$ is carried out under the condition that 
\begin{align}\label{l interval}
    -\frac{q}{2} + \frac{q\Delta \phi}{2\pi} \leq \ell \leq \frac{q}{2} + \frac{q\Delta \phi}{2\pi} .
\end{align}
 In particular, when ${\pm\frac{q}{2} + \frac{q\Delta \phi}{2\pi}}$ are integers, the associated terms in the summation must be adjusted with an additional factor of $1/2$. 
 As a result, we derive the following expression for $\mathcal{S}_s(r, q, \kappa) $
 %\begin{align}
 %     \mathcal{S}_s(r, q, \kappa) &= \frac{e^{\frac{is \Delta \phi}{2}}}{q}\sum_{\ell}(-1)^\ell  e^{-2 \pi  i \ell h_s} e^{r \cos  \left(\frac{2 \pi  \ell}{q}-\Delta \phi \right)} \nonumber\\
 %     &- \frac{e^{\frac{is \Delta \phi}{2}}}{\pi} \sum_{n=-\infty}^{\infty} e^{i \mu_{s}(n, \nu)\Delta \phi} \sin \left(\pi  |\mu_{s}(n, \nu)|\right)\times\nonumber\\ 
 %     &\times \int_0^{\infty }dy\,e^{-|\mu_{s}(n, \nu)|y - r\cosh (y)}
 %\end{align}
 \begin{align}
      &\mathcal{S}_s(r, q, \kappa) = \frac{e^{\frac{is \Delta \phi}{2}}}{q}\sum_{\ell}(-1)^\ell  e^{-2 \pi  i \ell h_s} e^{r \cos  \left(\frac{2 \pi  \ell}{q}-\Delta \phi \right)} \nonumber\\
      &- \frac{e^{\frac{is \Delta \phi}{2}}}{\pi} \int_{\infty}^{\infty} dx  e^{i q x \Delta \phi} \sum_{n=-\infty}^{\infty} F(n) \sin \left(q\pi  |x|\right)\times\nonumber\\ 
      &\times \int_0^{\infty }dy\,e^{-q|x|y - r\cosh (y)}.
 \end{align}
In the above expression, the function $F(n)$ is defined as
\begin{align}
F(n) = \delta\left[x - \left(n + \frac{1}{2} - h_s\right)\right],
\end{align}
and the delta function property $\int dx f(x) \delta(x - x_0)  = f(x_0)$ has been employed. Utilizing \eqref{Delta 1}, we can derive the following identity for the summation over $n$ in $F(n)$
\begin{align}
        \sum_{n=-\infty}^{\infty} F(n) =  \sum_{n=-\infty}^{\infty} e^{i2\pi n\left(x+h_s-\frac{1}{2}\right) } . 
\end{align}
%With this, we determine for $ \mathcal{S}_s(r, q, \kappa) $ that the following integral representation holds:
This helps us to present  $\mathcal{S}_s(r, q, \kappa)$ as follows
\begin{align}
      &\mathcal{S}_s(r, q, \kappa) = \frac{e^{\frac{is \Delta \phi}{2}}}{q}\sum_{\ell}(-1)^\ell  e^{-2 \pi  i \ell h_s} e^{r \cos  \left(\frac{2 \pi  \ell}{q}-\Delta \phi \right)} \nonumber\\
      &-\frac{e^{\frac{i s \Delta \phi}{2}}}{\pi} \sum_{n=-\infty}^{\infty}(-1)^n e^{i2\pi n h_s }  \int_0^{\infty }dy\,e^{- r\cosh (y)}\times \nonumber\\
      &\times \int_{-\infty}^{\infty} dx\,e^{i x (q \Delta \phi +2\pi n)}  \sin \left(q\pi  |x|\right)e^{-q|x|y}. 
 \end{align}
Evaluating the integral over the variable $x$, we obtain the following integral representation
\begin{align}
      \mathcal{S}_s(r, q, \kappa) &= \frac{e^{\frac{is \Delta \phi}{2}}}{q}\sum_{\ell}(-1)^\ell  e^{-2 \pi  i \ell h_s} e^{r \cos  \left(\frac{2 \pi  \ell}{q}-\Delta \phi \right)} \nonumber\\
      &- \frac{e^{\frac{i s \Delta \phi}{2}}}{\pi^2} \sum_{n=-\infty}^{\infty}(-1)^n e^{2\pi i n h_s } \times\nonumber\\
      & \times\int_0^{\infty }dy\,e^{- r\cosh (y)} M_{n,q}(\Delta \phi, y), 
 \end{align}
 where
 \begin{align}
     M_{n,q}(\Delta \phi, y) &= \frac{1}{2}\left[ \frac{n+\frac{q}{2}+ \frac{q\Delta \phi}{2\pi}}{\left(n+\frac{q}{2}+ \frac{q\Delta \phi}{2\pi} \right)^2 +\left(\frac{qy}{2\pi}\right)^2}\right.\nonumber\\
    &- \left. \frac{n-\frac{q}{2}+ \frac{q\Delta \phi}{2\pi} }{\left(n-\frac{q}{2}+ \frac{q\Delta \phi}{2\pi}\right)^2 +\left(\frac{qy}{2\pi}\right)^2}\right ] .
 \end{align}

With this expression at our disposal, we proceed to evaluate the integral provided in Eq. \eqref{Integral A}, conveniently changing the integration variable to $h_s$
\begin{align}\label{Integral B}
    &\mathcal{I}_s(r, q, \kappa) = e^{-\frac{is\Delta Z}{2\kappa} - \frac{\tau}{4\kappa^2}}\times \nonumber\\
    &\times \frac{q}{\kappa}\int_{-\infty}^{\infty}dh_s\,e^{-\frac{q^2\tau}{\kappa^2} h_s^2 + \frac{iq}{\kappa} \left(\Delta Z -\frac{is\tau}{\kappa} \right) h_s } \mathcal{S}_s(r, q, \kappa) .
\end{align}
Considering the first term of $\mathcal{S}_s(r, q, \kappa)$, we find
\begin{align}
    & e^{\frac{is \Delta \phi}{2} -\frac{is\Delta Z}{2\kappa} - \frac{\tau}{4\kappa^2}}  \sum_{\ell}(-1)^\ell  e^{r \cos  \left(\frac{2 \pi  \ell}{q}-\Delta \phi \right)}\times \nonumber\\
    &\times \frac{1}{\kappa}\int_{-\infty}^{\infty}dh_s\,e^{-\frac{q^2\tau}{\kappa^2} h_s^2 + \frac{iq}{\kappa} \left(\Delta Z -\frac{is\tau}{\kappa} - \bar{\kappa} \ell \right) h_s }\nonumber\\
    &=  \frac{e^{\frac{is \Delta \phi}{2}}}{q} \sqrt{\frac{\pi}{\tau}}\sum_{\ell}(-1)^\ell e^{-\frac{i\pi \ell s}{q}} e^{r \cos  \left(\frac{2 \pi  \ell}{q}-\Delta \phi \right)} e^{-\frac{\left(\Delta Z - \bar{\kappa} \ell \right)^2}{4\tau}} 
\end{align}
For the subsequent term in $\mathcal{S}_s(r, q, \kappa)$, we have
\begin{align}
    & \frac{1}{\pi^2} e^{\frac{is \Delta \phi}{2} -\frac{is\Delta Z}{2\kappa} - \frac{\tau}{4\kappa^2}} \sum_{n=-\infty}^{\infty}(-1)^n \times  \nonumber\\
    &\times \int_{0}^{\infty}dy\,e^{-r \cos(y)}  M_{n,q}(\Delta \phi, y)\times \nonumber\\
    &\times \frac{q}{\kappa}\int_{-\infty}^{\infty}dh_s\,e^{-\frac{q^2\tau}{\kappa^2} h_s^2 + \frac{iq}{\kappa} \left(\Delta Z -\frac{is\tau}{\kappa} + \bar{\kappa} n\right) h_s } \nonumber\\
    &=  \frac{e^{\frac{is \Delta \phi}{2}}}{\pi^2} \sqrt{\frac{\pi}{\tau}}\sum_{n=-\infty}^{\infty}(-1)^n  e^{\frac{i\pi n s}{q}} \, e^{-\frac{\left(\Delta Z +\bar{\kappa} n \right)^2}{4\tau}}  \times  \nonumber\\
    &\times \int_{0}^{\infty}dy\,e^{-r \cos(y)}  M_{n,q}(\Delta \phi, y) .
\end{align}

Accordingly, the resulting expression for the integral $\mathcal{I}_s(r, q, \kappa)$ is as follows
\begin{align}
 & \mathcal{I}_s(r, q, \kappa)  \nonumber\\
  &= \frac{e^{\frac{is \Delta \phi}{2}}}{q}\sqrt{\frac{\pi}{\tau}}\left[\sum_{\ell} (-1)^{\ell} e^{-\frac{ i\pi \ell s}{q} + r \cos\left(\frac{2\pi \ell }q -\Delta \phi\right) -\frac{(\Delta Z -\bar{\kappa} \ell )^2}{4\tau}}    \right.\nonumber\\
  &- \frac{q}{\pi^2} \sum_{n=-\infty}^{\infty} (-1)^{n} e^{\frac{ i\pi n s}{q}} \times \nonumber\\
  & \times \left.\int_{0}^{\infty} dy\,e^{- r \cosh(y) -\frac{(\Delta Z +\bar{\kappa} n )^2}{4\tau}} \mathcal{M}_{n,q}(\Delta \phi, y)  \right]. 
\end{align}
From the expression above, we obtain the following representation for the spinor heat kernel components $K_s$, as defined in {Eq.~\eqref{Appendix 1}}
\begin{align}
  &K_s  = \frac{q}{8\pi^2 \tau}e^{-\frac{\rho^2 + \rho'^2}{4\tau}-\tau m^2}\,\mathcal{I}_s(r, q, \kappa) \nonumber\\
  & =\frac{e^{-\tau m^2 +\frac{is \Delta \phi}{2}}}{(4\pi \tau)^{3/2} } \left[\sum_{\ell} (-1)^{\ell} e^{-\frac{ i\pi \ell s}{q}} e^{-\frac{\Delta \zeta_{\ell}^2}{4\tau}}    \right.\nonumber\\
&-\left. \frac{q}{\pi^2} \sum_{n=-\infty}^{\infty} (-1)^n e^{\frac{ i\pi n s}{q}}  \int_{0}^{\infty} dy \,e^{-\frac{\Delta \zeta_{n,y}^2}{4\tau}} M_{n,q}(\Delta \phi, y)  \right] , 
\end{align}
where the quantities $\Delta \zeta_{\ell}$ and $\Delta \zeta_{n,y}$ follow the definitions provided in {Eq.~\eqref{Delta Zeta}}.

\end{appendices}

\end{document}